\RequirePackage{rotating} 
\documentclass[fleqn,usenatbib]{mnras}

\usepackage{newtxtext,newtxmath}
\usepackage{rotating}

\usepackage[T1]{fontenc}

\DeclareRobustCommand{\VAN}[3]{#2}
\let\VANthebibliography\thebibliography
\def\thebibliography{\DeclareRobustCommand{\VAN}[3]{##3}\VANthebibliography}

\usepackage{graphicx}	
\usepackage{amsmath}	

\usepackage[export]{adjustbox}

\DeclareMathOperator*{\argmin}{arg\,min}

\title[DIA with PyTorch]{PyTorchDIA: A flexible, GPU-accelerated numerical approach to Difference Image Analysis}

\author[J. A. Hitchcock et al.]{
James A. Hitchcock $^{1}$\thanks{E-mail: jah36@st-andrews.ac.uk},
Markus Hundertmark $^{2}$,
Daniel Foreman-Mackey $^{3}$,
\newauthor
Etienne Bachelet $^{4}$,
Martin Dominik $^{1}$,
Rachel Street $^{4}$ and
Yiannis Tsapras $^{2}$ 
\\
$^{1}$ SUPA, School of Physics and Astronomy, University of St. Andrews, North Haugh, KY16 9SS, UK\\
$^{2}$ Astronomisches Rechen-Institut, Zentrum f{\"u}r Astronomie der Universit{\"a}t Heidelberg (ZAH), 69120 Heidelberg, Germany   \\
$^{3}$ Center for Computational Astrophysics, Flatiron Institute, Simons Foundation, 162 5th Ave, New York, NY 10010, USA\\
$^{4}$ Las Cumbres Observatory, 6740 Cortona Drive, Suite 102,93117 Goleta, CA, USA
}

\date{Accepted XXX. Received YYY; in original form ZZZ}

\pubyear{2020}

\begin{document}
\label{firstpage}
\pagerange{\pageref{firstpage}--\pageref{lastpage}}
\maketitle

\begin{abstract}
We present a GPU-accelerated numerical approach for fast kernel and differential background solutions. The model image proposed in the Bramich (2008) difference image analysis algorithm is analogous to a very simple Convolutional Neural Network (CNN), with a single convolutional filter (i.e. the kernel) and an added scalar bias (i.e. the differential background). Here, we do not solve for the discrete pixel array in the classical, analytical linear least-squares sense. Instead, by making use of PyTorch tensors (GPU compatible multi-dimensional matrices) and associated deep learning tools, we solve for the kernel via an inherently massively parallel optimisation. By casting the Difference Image Analysis (DIA) problem as a GPU-accelerated optimisation which utilises automatic differentiation tools, our algorithm is both flexible to the choice of scalar objective function, and can perform DIA on astronomical data sets at least an order of magnitude faster than its classical analogue. More generally, we demonstrate that tools developed for machine learning can be used to address generic data analysis and modelling problems.
\end{abstract}




\section{Introduction}\label{sec:introduction}

Difference Image Analysis (DIA) describes several astronomical image processing algorithms with the shared goal of delivering precise photometric measurements of variable astronomical sources. Given two images of the same scene acquired at different times, in a DIA framework, one aims to subtract one image from the other and recover a \textit{difference image} from which differential fluxes can be directly measured. This makes DIA a particularly effective approach to measuring the photometric variability of objects in crowded stellar fields -- such as microlensing campaigns e.g. \citep[][]{wozniak2000difference, bond2001real} and studies of globular clusters e.g. \citep[][]{bramich2011ccd, kains2012constraining, jaimes2013variable} -- where the blending of light from neighbouring sources cannot otherwise be easily disentangled.

Even for two perfectly aligned images from the same instrument, in order to produce a clean subtraction, the DIA algorithm of choice must model the inevitable changes in 1) the Point spread function (PSF) due to variations in seeing, telescope focus or tracking errors, 2) the Photometric scaling from differences in atmospheric transparency, exposure time or instrument throughput and 3) the Sky background caused by e.g. changes in the position and phase of the moon.\footnote{We note that there are other differences between images that cannot be captured by current DIA models, such as differential refraction or extinction, and we do not address these in this work.}

The differential change in PSF and photometric scaling can be found by inferring the \textit{convolution kernel}, which `blurs' the sharper of the two images to match the PSF in the other. \citet{alard1998method} (hereafter AL98) demonstrated that the kernel can be found using linear least-squares by decomposing it as a set of user-specified basis functions. Specifically, AL98 opted for Gaussian basis functions modified by low-order polynomials. A decade later, the linear least-squares solution was advanced by \citet{bramich2008new} (hereafter B08) who modelled the kernel as a highly flexible (albeit computationally more expensive) discrete pixel array, which is analogous to the AL98 algorithm, but with a choice of delta basis functions for the kernel model. Later advances included extending the DIA solution to a spatially varying kernel and differential background to model position-dependent variations in seeing, transparency and airmass across the field-of-view (FoV) in wide-field imaging data \citep[][]{alard2000image, bramich2013difference}.

In this analytical linear least-squares framework, a normal matrix for the linear system of equations is required. It is the construction of the normal matrix for these approaches which results in a computational bottleneck. For example, for a (square) pair of images each of size $n$, and a (square) kernel of size $m$ the construction of the normal matrix for the B08 approach -- which implements the same model for the kernel as our algorithm -- scales as $\mathcal{O}(n^2m^4)$ (i.e. the run-time increases with the square of the input image size and with the kernel size to the power of four).  These problems are exacerbated by the need to iteratively fit for the model parameters in the linear least-squares sense (as the problem is in fact non-linear, see Section \ref{sec:DIA}) and, typically, the normal matrix will need to be constructed about 3-4 times until convergence is reached. Attempts have been made to exploit symmetries in the problem (see Section 5.2 of \citet{bramich2013difference}) or bin pixels around the kernel edges to speed up this construction \citep{albrow2009difference}. In the era of wide FoV sky surveys such as the upcoming Legacy Survey of Space and Time (LSST) \citep{ivezic2019lsst}, or the ongoing Zwicky Transient Facility (ZTF) \citep{bellm2018zwicky} etc., which aim to deliver prompt alerts of transient astronomical phenomena detected through image subtraction, many current popular approaches make real-time event discovery impractical. This has driven some recent advances in the DIA literature, most notably the ZOGY algorithm \citep{zackay2016proper}, and even a machine-learning approach \citep{sedaghat2018effective}.

Image processing tasks -- where some common operation is performed on very many pixels -- are inherently massively parallel. Mild to substantial computational speed-up by parallelising the construction of the normal matrix in classical DIA algorithms with GPUs has been demonstrated in the literature \citep[for example,][]{hartung2012gpu, li2013gpu, zhao2013accelerating}. Adoption by the larger astronomical community however has been slow.

The main barrier to adoption for most astronomers is likely the lack of a working knowledge of CUDA, NVIDIA's parallel computing platform, which is required to develop GPU-accelerated applications on supported devices. In astronomy, the Python programming language has firmly established itself as a favourite of the majority of the community, and most importantly, will be the primary user-interface language for the next generation of astronomical data management, processing and analysis systems \citep{perkel2018jupyter}. The appropriateness of DIA image models can be highly data set dependent, and tuning by individual researchers to meet their science goals within a Pythonic framework is to be expected. The work presented here is one of a few recent attempts\footnote{Including approaches attempting to parallise the construction of the B08 normal matrix \citep{michaeldalbrow_2017_268049}} to address the paired issues of performance and accessibility.

In this paper, we present a novel Pythonic implementation of an alternative route to the B08 solution, without the need for constructing the computationally expensive normal matrix. Conceptually, our approach is unique in that we model the kernel as if it were the convolutional filter of a very simple convolutional neural network (CNN), which can be solved for efficiently with GPU-accelerated optimisation tools originally developed for deep learning applications. As we will describe, the machine-learning framework which we use to approach this problem also equips us with powerful modelling tools, and frees us from a number of restrictive assumptions inherent to classical approaches. Our implementation is also unique amongst GPU-accelerated DIA algorithms for being written \textit{entirely} with standard Python packages.

We begin Section \ref{sec:problem} with an introduction of the basic image model used in our DIA implementation. We then outline how PyTorch could be well suited to addressing existing astronomical image modelling and data processing challenges in general, before describing the details of our own DIA implementation, `PyTorchDIA'. We quantify the model fit quality and photometric accuracy of PyTorchDIA with tests on both synthetic and real images in Sections \ref{sec:simulatedimagetests} and \ref{sec:realimage_tests}, comparing it directly to the performance of its classical DIA analogue, the B08 algorithm. In Section \ref{sec:speed_test}, we compare the speed of our GPU-accelerated numerical solution against a fast Cython implementation of the B08 algorithm\footnote{The kernel solution method heavily borrows from the relevant section of the pyDANDIA microlensing reduction pipeline, \url{https://github.com/pyDANDIA}.} used in the ROME/REA project \citep{tsapras2019rome}, and explore how our algorithm scales with image and kernel size. We summarise our conclusions in Section \ref{sec:conclusions}.

\section{Problem Formulation}\label{sec:problem}

In this section we outline the DIA problem, and provide a motivating overview for using PyTorch as a tool to address image modelling challenges -- of which DIA is just one example -- and describe the details of our DIA implementation.

\subsection{Difference Image Analysis}\label{sec:DIA}

Given a reference image with pixels $R_{ij}$, ideally of excellent spatial resolution and high signal-to-noise, and a target image with pixels $I_{ij}$, of the same scene taken at some other epoch and aligned on the same pixel grid, the model image which represents $I_{ij}$ is given by
\begin{equation}
    M_{ij} = [R \otimes K]_{ij} + B_{ij} \;,
    \label{eq:model_image}
\end{equation}
and so the challenge is to find the fit for an accurate kernel, $K$, and differential background $B_{ij}$.

Following B08, by representing the kernel as a discrete array $K_{lm}$ containing a total of $N_K$ pixels, and including an additive scalar differential background, $B_0$, we can re-write equation \ref{eq:model_image} as
\begin{equation}
    M_{ij} = \sum_{lm} K_{lm}R_{(i+l)(j+m)} + B_0 \;.
    \label{eq:model_image_delta_basis}
\end{equation}
The photometric scale factor -- which encodes any differences in atmospheric transparency and/or exposure time between the images -- is simply the sum of the kernel pixels,
\begin{equation}
    P = \sum_{lm} K_{lm} \;.
    \label{eq:photometric_scale_factor}
\end{equation}
Assuming that the pixel values of the target image, $I_{ij}$, are independently drawn from normal distributions $\mathcal{N}(M_{ij}, \sigma^{2}_{ij})$ -- where $M_{ij}$ is paramaterised by the vector $\boldsymbol{\theta}=[K_{lm}, B_0]$ (see Section \ref{sec:DIAasoptim}) --  the negative log-likelihood function for the target image takes the form
\begin{equation}
    -\textrm{ln}\;\textrm{p}(I_{ij} | \boldsymbol{\theta}) = \frac{1}{2}\chi^2\ + \sum_{ij} \ln\;\sigma_{ij} + \frac{N_\mathrm{data}}{2} \textrm{ln}(2\mathrm{\uppi})  \;,
    \label{eq:loglikelihood}
\end{equation}
where $\sigma_{ij}$ are the pixel uncertainties, $N_{\mathrm{data}}$ is the number of pixels in the target image, and the $\chi^2$ is equal to
\begin{equation}
    \chi^2 = \sum_{ij}\left(\frac{I_{ij} - M_{ij}}{\sigma_{ij}} \right)^2 \;.
    \label{eq:chi_squared}
\end{equation}

It is important to note that $N_{\mathrm{data}}$ is \textit{not} equal to the total number of pixels in the target image, as the convolution operation is undefined for pixels within half a kernel's width from the target image edges i.e. for a kernel of size $(2n + 1) \times (2n + 1)$ and (square) reference and target images of size $N \times N$, $N_{\mathrm{data}} = (N - 2n)^2$.

The $\sigma_{ij}$ pixel uncertainties are dependent on the image model $M_{ij}$, and so fitting this model to the target image $I_{ij}$ is therefore a non-linear optimisation task. Linear least-squares approaches like AL98 and B08 must then approach this iteratively, by minimising the $\chi^2$ with fixed estimates for $\sigma_{ij}$. After the first estimate of $M_{ij}$ is acquired, the $\sigma_{ij}$ are computed, and then plugged into the $\chi^2$ for the next iteration. This process continues for at least 3 iterations, until some convergence condition is met.

In this work, we use both simulated CCD images and real Electron Multiplying CCD (EMCCD) images, acquired on the Danish 1.54m (DK154) Lucky Imaging camera \citep{skottfelt2015two} to verify our implementation, each of which requires a different noise model. In what follows, we ignore the noise contributions from the reference image, since these are negligible in the experiments performed in this work.  \footnote{Our interest in testing the performance of our DIA algorithm on DK154 EMCCD images relates to our ongoing microlensing follow-up campaign, which is the main science project of the MiNDSTEp consortium, \url{http://www.mindstep-science.org/}}.

For CCD images, we adopt a noise model for the $\sigma_{ij}$ pixel uncertainties of $M_{ij}$ as
\begin{equation}
    \sigma_{ij}^{2} = \frac{\sigma_{0}^{2}}{F_{ij}^{2}} + \frac{M_{ij}}{G\;F_{ij}} \;
    \label{eq:CCDNM}
\end{equation}
where $\sigma_0$ is the read noise (ADU), $G$ is the detector gain ($\textrm{e}^- / \textrm{ADU}$), and $F_{ij}$ is the master flat field.

The EMCCD images are constructed from typically thousands of shift-and-added sub-second exposures. The electron multiplying gain reduces the read out noise to negligible levels, but the cascade amplification process effectively doubles the variance of the photon noise. For the real EMCCD images used in this work, we adopt a noise model of the form
\begin{equation}
    \sigma_{ij}^{2} = E \frac{M_{ij}}{G_{\textrm{Total}}\;F_{ij}}, \;
    \label{eq:EMCCDNM}
\end{equation}
where $G_{\textrm{Total}}$ represents the combined gain ($\text{e}_{\text{phot}}^{-}$/ADU) of the DK154 Lucky Imaging camera, which is calculated as the ratio of the CCD gain of 25.8 ($\text{e}_{\text{EM}}^{-}$/ADU) over the electron-multiplying (EM) gain of 300 ($\text{e}_{\text{EM}}^{-}$/$\text{e}_{\text{phot}}^{-}$). Following \citet{harpsoe2012high}, we differentiate between electrons before and after the cascade amplification with the notation $\text{e}_{\text{phot}}^{-}$ and $\text{e}_{\text{EM}}^{-}$ respectively. $E$ represents the `excess noise factor', and accounts for the probabilistic nature of the cascade amplification as the EMCCD is read out, and is set to be $E=2$. EMCCD images are flat corrected in the same way as conventional CCD images, and as in Equation \ref{eq:CCDNM}, $F_{ij}$ is the master flat field.

\subsection{Astronomical Image Processing with PyTorch}

The model image of Equation \ref{eq:model_image} is a convolution with some added scalar constant. In the computer science literature, this is \textit{exactly} analogous to an extremely simple convolution neural network (CNN), with a single input and output related by a single convolutional filter, with some additional scalar bias added to the output. Efficient solutions for the `weights' (the kernel) and `bias' (background term) to this convolution operation can be implemented within the Python package, PyTorch \citep{paszke2019pytorch}.

PyTorch is a popular open source machine learning framework. Its constant development is motivated by the impressive advances in deep learning (for an overview see e.g. \citep[][]{lecun2015deep, goodfellow2016deep}), in which CNNs have played an important role in processing images. Specifically, this package supports CUDA-enabled GPU acceleration and automatic differentiation to perform efficient optimisations. One key motivation behind these developments is the \textit{training} of complex CNNs, consisting of thousands, to hundreds of thousands of parameters. The 2-dimensional convolution (as in Equation \ref{eq:model_image}) is \textit{the} core processing operation in these networks, and is straightforward to implement in PyTorch's modelling architecture. Indeed, many useful image models in astronomy can be written as a convolution, and the tools outlined in this work are therefore broadly applicable to many astronomical image processing problems.

We stress that although PyTorch's powerful tools were designed for machine learning, they can be turned to generic data analysis and modelling problems, as shown in this work, of which image models are just a subset. In particular, efficient computation of gradients via automatic differentiation frees the user from having to manually recompute gradients if the parameterisation of the model is changed. This flexibility allows models written in PyTorch to be easily tuned to meet the variety of science goals arising from a diversity of data sets. In astronomical image processing in particular, this flexibility combined with GPU acceleration could make PyTorch a valuable tool for addressing challenges associated with both model complexity and data volume. As PyTorch is Pythonic, these implementations could be easily integrated into existing Python stacks.

\subsection{Difference Image Analysis as an Optimisation}\label{sec:DIAasoptim}

The crucial advance in AL98 was to formulate DIA as a linear least-squares problem, and several efficient algorithms for solving these problems exist \citep{golub1996matrix}. The computational bottleneck associated with this approach is the construction of the normal matrix (see Section \ref{sec:introduction}). Models in PyTorch however are generally fit with a numerical optimisation procedure -- making use of automatic differentiation -- which we outline here.

For a given target image with Gaussian noise contributions, $I_{ij}$, and current estimates for the kernel $K_{lm}$ and differential background $B_0$ which transform the reference image, $R_{ij}$, we define our vector of weights as $\boldsymbol{\theta} = [K_{lm}, B_0]$. Ignoring the irrelevant normalisation constant, the maximum-likelihood-estimate (MLE) for $\boldsymbol{\theta}$ can be found by minimising the overall \textit{loss} (or negative log-likelihood Cf. Equation \ref{eq:loglikelihood}),
\begin{align}
\begin{split}
&\argmin_{\boldsymbol{\theta} = [K_{lm}, B_0]}  \mathcal{L}_0(\boldsymbol{\theta}) = \\
&= \argmin_{\boldsymbol{\theta}} \; \left[
\frac{1}{2}\sum_{ij}  \left(\frac{I_{ij} - M_{ij}(\boldsymbol{\theta})}{\sigma_{ij}(\boldsymbol{\theta})}\right)^2 + \sum_{ij} \ln\;\sigma_{ij}(\boldsymbol{\theta})\right] \;.
\end{split}
\label{eq:loss}
\end{align}

We note here that both the image model and noise model are both functions of $\boldsymbol{\theta}$. We drop the notation explicitly showing the dependence of $M_{ij}$ and $\sigma_{ij}$ on $\boldsymbol{\theta}$ from the rest of the manuscript.

Equation \ref{eq:loss} can be solved by a process of (steepest) gradient descent. At each iteration $t$ in the optimisation, we use the update rule
\begin{equation}
    \boldsymbol{\theta}^{(t+1)} \; \longleftarrow \; \boldsymbol{\theta}^{(t)} - \alpha^{(t)}\nabla_{\boldsymbol{\theta}^{(t)}}\mathcal{L}_0(\boldsymbol{\theta}^{(t)}) \;,
    \label{eq:update}
\end{equation}
where $\alpha^{t}$ is the \textit{learning rate} (or step-size) at each iteration. PyTorch includes implementations of several more sophisticated gradient descent algorithms, which can compute \textit{adaptive} learning rates for \textit{each} parameter (see \citealt{ruder2016overview} for a good overview). In our implementation, we use the Adaptive Moment Estimation (Adam) algorithm , which computes learning rates for the kernel pixels and the background parameters from estimates of the first and second moments of their gradients at each step \citep{kingma2014adam}. Adam has been empirically demonstrated to work well on non-convex optimisation problems, is computationally efficient, and the hyper-parameters are both intuitive and require little tuning from the astronomer (see Section \ref{sec:engineering} for an overview of the `engineering' aspects of the optimisation).

With an excellent choice of learning rate, solutions via steepest descent when accelerated on the GPU are lightning fast (see Section \ref{sec:speed_test}), but this approach in general can be slow. Solving this problem on the CPU -- and therefore forgoing the massive inherent parallelism otherwise exploited in Equation \ref{eq:update} -- would result in a substantial performance hit. This problem is particularly severe when close to the minimum of the loss function, where the gradients become increasingly shallow and the gradient steps become correspondingly smaller. An effective solution to this problem is to make use of quasi-Newton optimisation methods which approximate the curvature of the loss surface.

These approaches converge extremely quickly where the loss surface can be modeled quadratically, and use an approximation of the inverse of the Hessian at each $t$ iteration, $B(\boldsymbol{\theta})^{(t)}$, to condition the search direction, giving the update rule
\begin{equation}
    \boldsymbol{\theta}^{(t+1)} \; \longleftarrow \; \boldsymbol{\theta}^{(t)} - \alpha^{(t)}B^{-1}(\boldsymbol{\theta})^{(t)}\nabla_{\boldsymbol{\theta}^{(t)}}\mathcal{L}_0(\boldsymbol{\theta}^{t}) \;.
    \label{eq:newtonupdate}
\end{equation}

For this convex optimisation, this is a very good assumption when close to the minimum. Newton methods in general are sensitive to bad parameter initialisation (i.e.\ when we're initially far from the minimum),
and so we advocate for optimising via steepest descent steps -- as in Equation \ref{eq:update} -- to first get close to the minimum, before converging with the more memory intensive quasi-Newton procedure once the relative change in the loss between any two optimisation steps falls below a user-specified threshold. Specifically, we use a L-BFGS method (see Chapter 7.2 of \citealt{nocedal2006numerical}) which is available as an in-built algorithm in PyTorch\footnote{\url{https://pytorch.org/docs/stable/optim.html}}.

\subsection{Robust loss function}

In general, the assumption that all the pixel values in the target image are drawn from $\mathcal{N}(M_{ij}, \sigma_{ij}^2)$ will be violated for real images. Instrumental defects, cosmic ray hits and variable or transient sources etc. will all arise as outlying pixel values. Our Gaussian loss function (Equation \ref{eq:loglikelihood}) will be badly affected by these outliers, as it minimises the squared difference between the model and the data. The standard approach to outlier rejection in astronomy is to iteratively remove these `bad' pixels by \textit{sigma-clipping}, and B08 use this approach to mitigate the impact of outliers to the least-squares solution. In addition to being very sensitive to the accuracy of the adopted noise model, this \textit{procedure} does not explicitly penalise the rejection of data, nor is it necessarily clear how many iterations should be performed. In the context of classical DIA algorithms, these iterations incur an extra computational expense, as a new normal matrix must be constructed each time.

\begin{figure}
    \centering
    \includegraphics[width=\columnwidth]{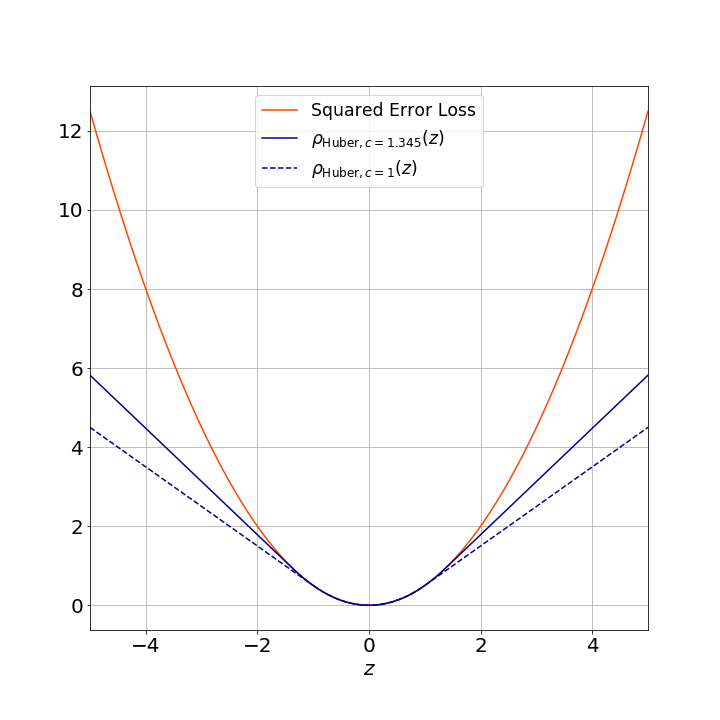}
    \caption{A comparison of squared error loss against Huber loss for different values of $c$.}
    \label{fig:huberloss}
\end{figure}

As we are not restricted to standard least-squares minimisation with our optimisation procedure, we have the freedom to choose \textit{robust} alternatives to Equation \ref{eq:loss} which are less sensitive to outlying values. \citet{huber1992robust} identified a family of wide-tailed univariate distributions. The `$\epsilon$-contaminated' Gaussian model in particular corresponds to a unimodal symmetric distribution which resembles a Gaussian for central values and a Laplacian in the tails. The probability density function (up to a normalising constant) with location and scale parameters $\mu$ and $\sigma$ has the form
\begin{equation}
    f(x) \propto  \frac{1}{\sigma} \exp \left[-\rho_{\text{Huber}, c}\left(\frac{x - \mu}{\sigma}\right)\right] \;,
    \label{eq:huber_pdf}
\end{equation}
Substituting $z = (x - \mu) / \sigma$, $\rho_{\text{Huber}, c}(z)$ is the loss function, which is defined as
\begin{equation}
    \rho_{\text{Huber}, c}(z) =
    \begin{cases}
    \frac{1}{2}z^2, & \text{for } |z|\leq c \\
    c(|z| - \frac{1}{2}c), & \text{for } |z| > c \;.
    \end{cases}
    \label{eq:huber_loss}
\end{equation}
We see that $\rho_{\text{Huber}, c}(z)$ computes the squared error between the model and data for small residuals, and the absolute deviation for outliers above some user-specified threshold, $c$. This threshold defines the switch between quadratic and linear treatment of the error (see Figure \ref{fig:huberloss}), and \citet{huber2009robust} recommend $c=1.345$ as a suitable value for enforcing robustness while retaining reasonable efficiency for normally distributed data.

Using the same notation as in Section \ref{sec:DIA}, if we assume the pixel values in the target image $I_{ij}$ are drawn from the distribution in Equation \ref{eq:huber_pdf}, the negative log-likelihood function takes the general form
\begin{equation}
 -\ln p(I_{ij}|\boldsymbol{\theta}) =
 \sum_{ij}\rho_{\text{Huber}, c} \left(\frac{I_{ij} - M_{ij}}{\sigma_{ij}}\right) + \sum_{ij} \ln \sigma_{ij} + Q \;,
 \label{eq:huber_nll}
\end{equation}

where $Q$ is the normalising constant,

\begin{equation}
    Q = N \ln \left(\sqrt{2\mathrm{\uppi}} \, \textrm{erf}\left(\frac{c}{\sqrt{2}}\right) + 
        \frac{2\, \textrm{exp}\left[\frac{-c^2}{2}\right]}{c}\right) \;.
\end{equation}\label{eq:huber_constant}

Note that when $c$ becomes large, $Q = (N / 2) \times \ln{2\mathrm{\uppi}}$, which is equivalent to the normalisation constant of a Gaussian log-likelihood (Cf. Equation \ref{eq:loglikelihood}).  
 
%
%
%
%

As our implementation makes use of automatic differentiation, it is straightforward for the user to experiment with loss functions, as they are freed from having to manually recompute gradients. We highlight the Huber loss in particular as this enforces robustness without sacrificing performance (see Section \ref{sec:emccdspeed}), but any number of wide-tailed distributions (e.g. Student's t distribution) could be used as drop-in replacements.

\subsection{Uncertainty estimation - Observed Fisher Information}

The central limit theorem states that than any well-behaved likelihood function approaches a Gaussian near its maximum. The \textit{Fisher Information Matrix} (FIM) is a measure of the curvature of the likelihood function with respect to the model parameters -- intuitively, this can be thought of as a `sensitivity' -- and its inverse provides a lower bound on the asymptotic variance of the Maximum Likelihood Estimate (MLE).

The observed FIM, $\mathrm{\boldsymbol{F}}(\boldsymbol{\theta})$ for our $N_K + 1$ parameters is the $(N_K + 1,\;N_K +1)$ matrix containing the entries
\begin{equation}
    F_{pq} =
    \frac{\partial^2}{\partial\theta_p\partial\theta_q}\textrm{ln}\;\textrm{p}(I_{ij} | \boldsymbol{\theta})\;,\; 1\leq p,\; q\leq N_K + 1 \;,
    \label{eq:FIM}
\end{equation}
where $\textrm{ln}\;\textrm{p}(I_{ij} | \boldsymbol{\theta})$ is the log-likelihood (Equation \ref{eq:loglikelihood}).

The inverse of $\mathrm{\boldsymbol{F}}(\boldsymbol{\theta})$ evaluated at the MLE for $\boldsymbol{\theta}$ (i.e. $\boldsymbol{\hat{\theta}}_{\textrm{MLE}}$)  can then be used as an estimate for the covariance matrix
\begin{equation}
    \hat{\boldsymbol{\Sigma}} =  \textrm{Cov}(\boldsymbol{\hat{\theta}}_{\textrm{MLE}}) = [\mathrm{\boldsymbol{F}}(\boldsymbol{\hat{\theta}}_{\textrm{MLE}})]^{-1}\;,
    \label{eq:covmatrix}
\end{equation}
which provides an estimate of the uncertainties on the model parameters by taking the square roots of the diagonal elements of $\textrm{Cov}(\boldsymbol{\hat{\theta}}_{\textrm{MLE}})$. Fisher information provides us with the limiting precision with which model parameters can be estimated for any given data set i.e. the subsequent error bars cannot be smaller \citep{heavens2009statistical}. Formally, the \textit{Cram\'er-Rao} inequality states that the uncertainty on some parameter $\theta_p$ is given by

\begin{equation}
    \Delta\theta_p \geq (F^{-1})_{pp}^{1/2}\;.
    \label{eq:deltatheta}
\end{equation}

This method is subject to two assumptions 1) the likelihood function correctly describes the data generating process (i.e. the error distribution of the measurements is correctly described by the likelihood), and 2) the likelihood really is approximately Gaussian at the MLE. In practice, both of these assumptions will likely be violated, and in these situations, the uncertainty estimates by this approach can be severely underestimated. Given this, \citet{andrae2010error} strongly recommend to test this assumption by checking the validity of $\hat{\boldsymbol{\Sigma}}$ . In general, a valid covariance matrix $\hat{\boldsymbol{\Sigma}}$ must be positive definite (i.e. for any non-zero vector $\boldsymbol{x}$, $\boldsymbol{x}^{T}.\hat{\boldsymbol{\Sigma}}.\boldsymbol{x} > 0$). If either the determinant of $\hat{\boldsymbol{\Sigma}}$ is negative, or (after diagonalising the matrix) any eigenvalue is found to be negative or zero, then $\hat{\boldsymbol{\Sigma}}$ is not valid. We include both these tests in our code release, and a warning flag is raised if any is failed.

\subsection{Extension to a Spatially Varying Background}

Within the PyTorch architecture, a spatially varying background can easily modelled by replacing $B_{ij}$ in Equations \ref{eq:model_image} and \ref{eq:loss} with a linear combination of functions of $x$ and $y$. We adopt a polynomial model of some user-specified degree $d$,

\begin{equation}
    B(x, y) = \sum_{m=0}^{d}\sum_{n=0}^{d-m}  b_{mn}\, \eta(x)^{m} \xi(y)^{n} \;
    \label{eq:polybackground}
\end{equation}

where $b_{mn}$ are the polynomial coefficients to be inferred, and $\eta(x)$ and $\xi(y)$ are the normalised spatial coordinates,
\begin{equation}
    \begin{split}
    \eta(x) = (x - x_c)/N_x\;, \\
    \xi(y) = (y - y_c)/N_y\;,
    \end{split}
    \label{eq:normcoords}
\end{equation}
which result from a Taylor expansion of coordinates $(x, y)$ about the image centre $(x_c, y_c)$, for an image of $N_x \times N_y$ pixels. This choice of coordinates is recommended by \citet{bramich2013difference}, and has the effect of (1) improving the orthogonality of the spatial polynomial terms, and (2) preventing the coefficients of the higher order polynomial terms from being pushed towards zero.

\subsection{Regularising the kernel pixels}

As noted by \citet{becker2012regularization}, while kernels modeled as a discrete array of pixels are very flexible, the consequent fidelity with the data can result in significant overfitting. To guard against excessively noisy kernels, we provide the option to regularise the loss with the addition of a penalty term. Following the notation in \citet{bramich2016difference} (hereafter B16) -- where the strength of the regularisation is controlled by the parameter $\lambda$, which must be tuned empirically -- the scalar objective function to minimise now takes the form,
\begin{equation}
    \argmin_{\boldsymbol{\theta}} \mathcal{L}(\boldsymbol{\theta}) = \argmin_{\boldsymbol{\theta}}\left[\mathcal{L}_0(\boldsymbol{\theta}) + \lambda N_{\textrm{data}} \boldsymbol{\theta}^{\textrm{T}}\boldsymbol{L}^{\textrm{T}}\boldsymbol{L}\boldsymbol{\theta} \right]\;,
    \label{eq:regularised_loss}
\end{equation}
where $N_{\textrm{data}}$ is the number of target image pixels,  $\boldsymbol{\theta}$ is the vector of parameters to optimise, $\mathcal{L}_0(\boldsymbol{\theta})$ is the loss function, and $\boldsymbol{L}$ is the symmetric and positive-semidefinite $(N_K + 1) \times (N_K + 1)$ \textit{Laplacian matrix}, which represents the connectivity graph of the set of kernel pixels, and has elements
\begin{equation}
  L_{uv} =
    \begin{cases}
      N_{\textrm{adj},u}, & \text{for $v=u \leq N_K$, and $N_{\textrm{adj},u}$ is the number of }\\
      & \text{kernel pixels adjacent to the kernel pixel}\\
      & \text{corresponding to $u$}\\
      -1, & \text{for $u \leq N_K, v \leq N_K,  v \neq u$, and $u$ and $v$ are}\\
      & \text{adjacent kernel pixels.}\\
      0, & \text{otherwise.}
    \end{cases}
    \label{eq:laplacian}
\end{equation}

This penalty term is derived from an approximation of the second derivative of the kernel surface \citep{becker2012regularization}. Intuitively, it favours compact kernels, where adjacent kernel pixels should not vary too sharply. The optimal value of $\lambda$ should, ideally, be tuned for each image, and B16 found optimal values could be anywhere in the range $\lambda = 0.1 - 100$.

By regularising the kernel weights, we are in effect introducing a Bayesian prior, which would then transform our solution into a \textit{maximum a posteriori} (MAP) optimisation. In our experiments in the following sections, we restrict ourselves to the MLE solution by simply minimising either Equation \ref{eq:loss} or \ref{eq:huber_nll}.

\subsection{General purpose computing on GPUs}

For this work, we ran our PyTorchDIA implementation on two different NVIDIA GPUs. The computations associated with the results presented in Section \ref{sec:simulatedimagetests} and \ref{sec:realimage_tests} were performed on a GeForce GTX 1050 on a local machine, and the speed tests in Section \ref{sec:speed_test} were run on a Tesla K80 on Google's Colab\footnote{\url{https://colab.research.google.com}} service, a free-to-use cloud-based computation environment. There are a couple of important details worth describing on using these devices for scientific computing.

Firstly, while modern CPUs are able to handle computations on 64-bit floating point numbers efficiently, such operations are either not supported on GPUs, or are associated with a significant reduction in performance \citep{goddeke2005accelerating}. It is for this reason that 32-bit precision -- or increasingly commonly, 16-bit precision -- is used in deep learning. For this problem, we have found computations typically take at least $2-5$ times longer for fiducial image sizes using 64-bit precision with a Tesla K80 \footnote{Tesla cards are designed to perform fast computations at higher precision than most GPU models, and a slowdown of $\sim 2-5$ at 64-bit would still outperform some state-of-the-art classical DIA methods (see Section \ref{sec:speed_test}). Most NVIDIA models however (including GeForce cards), excel at 32-bit and lower precisions, but would suffer a more severe slowdown at 64-bit.}. For memory intensive optimisers such as L-BFGS, the hit to performance will be even worse. For this reason, for the results in this work we use 32-bit floating point precision for our PyTorchDIA implementation. This speed-up comes at the expense of some accuracy, but we show this difference to be small in Sections \ref{sec:simulatedimagetests} and \ref{sec:realimage_tests}.

Convolution computations on NVIDIA GPUs can be accelerated by highly-tuned algorithms available in specialised libraries. One such NVIDIA library, cuDNN, specialises in efficient computations with small kernels (i.e. those in the regime of DIA). These algorithms are selected by cuDNN heuristically, and not all are deterministic. As such, for an identical pair of images, the number of computations used to infer the convolution kernel may be different on different runs. Consequently, the optimiser may converge at a slightly different point in the parameter space. We explore the speed-up associated with enabling cuDNN tools in Section \ref{sec:speed_test}.

Finally, while not explored in this work, we highlight \textit{mixed precision} `training' as a technique which our approach could benefit from. PyTorch now supports Automatic Mixed Precision (AMP) training, in which 32-bit data can be automatically cast to half precision for some types of computationally expensive operations on the GPU\footnote{\url{https://pytorch.org/docs/stable/notes/amp_examples.html}}. By appropriately scaling the loss during the training/optimisation, AMP prevents \textit{underflow} that would otherwise cause gradients to drop to 0 at half precision, and can achieve the same accuracy as training only with 32-bit precision with significant speed-up, depending on the GPU architecture and model design \citep{micikevicius2017mixed}. Some recent NVIDIA GPUs now include \textit{Tensor cores}, which are specifically designed to perform highly optimised 16-bit matrix multiplications. Consequently, cuDNN convolution algorithms such as `GEMM' are particularly well suited to benefit from this technique \citep{jorda2019performance}.

\subsection{Optimisation as an engineering problem}\label{sec:engineering}

As we fit for the convolution kernel via an optimisation, hardware alone will not determine the solution time. There is an `engineering' aspect to optimisation that the user should be aware of. Here, we summarise three key choices that the user must make: 1) Parameter initialisation; 2) The learning rate (or step size) and 3) The convergence condition. We also provide an overview of (and justification for) the specific choices made when generating the results presented in this manuscript.


Throughout this work, we \textit{always} background subtract the reference image, and so we initially set the differential background parameter, $B_0$, to be equal to the median pixel value of the given target image. For the fit quality and photometric accuracy tests in Sections \ref{sec:simulatedimagetests} and \ref{sec:realimage_tests}, we make no assumptions about the shape of the kernel at initialisation. We set all kernel pixels to have the same value, with the only condition being that they sum to 1 (i.e. each pixel is initialised as $1/N_K$). In Section \ref{sec:speed_test}, we experiment with initialising the kernel as a symmetric Gaussian, with a shape parameter judiciously set with knowledge of either the known or measured differences in the PSFs of the reference and target image pair.

The Adam algorithm used at the start of the optimisation -- with which we perform steepest gradient descent -- allows us the freedom to set individual learning rates for our model parameters, which will be adaptively tuned (see Section \ref{sec:DIAasoptim}). For all tests in this work, we set the learning rate of the Adam optimizer for the pixels in the kernel at 0.001, as recommended in \citet{kingma2014adam}. We have found that it is advantageous to use a fairly high learning rate for $B_0$, and we set this to either 1, 10 or even 100, dependent on the data set. There is a strong anticorrelation between $P$ and $B_0$, and quickly finding the approximate photometric scaling between the two images allows us to disentangle this offset from the inference of the spatial differences in the images (associated with the different PSFs), which is encoded in the shape (and not the `scale') of the convolution kernel. The learning rate of the L-BFGS optimiser is always set to 1.

We use the same convergence condition for all tests in this work. This decision was made on the basis of the model performance and photometric accuracy metrics in Sections \ref{sec:simulatedimagetests} and \ref{sec:realimage_tests}
, and should balance a satisfactory model accuracy with the time taken to converge. If between any two steepest descent updates the relative change in the loss is less than $10^{-6}$ (i.e. we are getting close to the minimum) a L-BFGS optimiser takes over. For each subsequent quasi-Newton update, the L-BFGS function evaluation routine terminates when either the change in loss between evaluations reaches the limit of our numerical precision -- which corresponds to a relative change in loss in the range $10^{-8} - 10^{-7}$ between the last two optimisations steps -- or a first order optimality condition is met, such that the gradient of the scalar objective function to be minimised with respect to each model parameter is less than $10^{-7}$.

\subsection{The Algorithm}\label{sec:algorithm}

Our DIA algorithm is as follows. Decisions to be made by the user/human are in italics.  `tol' is the user-specified relative change in the loss between successive steepest descent (SD) updates, at which point the optimisation routine switches from SD to L-BFGS.

\begin{itemize}
    \item \textit{Choose the square kernel dimensions (must be odd), and whether to use a scalar, or polynomial fit of degree $d$, for the differential background}
    \item \textit{Set Adam's per-parameter learning rates and `tol'}
    \item \textit{Set the convergence condition}
    \item \textit{Initialise $\boldsymbol{\theta}$} 
    \item Begin minimising the chosen scalar objective function (Equations \ref{eq:loglikelihood} or \ref{eq:huber_nll}, with or without the optional penalty term (Equation \ref{eq:regularised_loss})) with steepest gradient descent (Equation\ref{eq:update})
    \item At each iteration \textbf{if} the relative change in the loss is less than `tol', switch to a quasi-Newton update (Equation \ref{eq:newtonupdate}) and \textbf{continue} until convergence condition satisfied.
\end{itemize}

\section{Simulated Image tests}\label{sec:simulatedimagetests}

Artificially generated images can be used to assess the accuracy of our algorithm. In this section, we compare the fit quality and photometric accuracy of our numerical, GPU-accelerated algorithm against an implementation of the analytic B08 algorithm with a data set of 71569 synthetic images. 

\subsection{Generating Artifical Images}\label{sec:genimages}

We base our image generation procedure on Section 5.1 of B16. The results from the simulated image tests in that work were shown to closely agree with similar tests using real CCD data. Specifically, we generate images similar to those in the `S10' set of that work, wherein the reference image of each image pair has 10 times less variance than the target.  

We first generate a noiseless reference, $R_{\textrm{noiseless}, ij}$, of size $142 \times 142$ pixels as follows.

(i) We generate a normalised two-dimensional symmetric Gaussian PSF for this image, parameterised by a standard deviation, $\phi_{R}$, drawn from the continuous uniform distribution $U \sim (0.5, 2.5)$.

(ii) We populate the $142\times142$ pixel array template of $R_{\textrm{noiseless}, ij}$ with $N_{\textrm{stars}}$, whose lg density $\lg \rho_{\textrm{stars}}$ per $100\times 100$ pixel region, is drawn from the uniform distribution $U \sim (0, 3)$. The fractional pixel coordinates for each star are uniformly drawn between the image axis lengths.

(iii) For each of the $N_{\textrm{stars}}$, we draw a value of $\mathcal{F}^{-3/2}$ from the uniform distribution $U \sim (10^{-9}, 10^{-9/2})$, where $\mathcal{F}$ is a given star's flux (ADU). This flux distribution is a good approximation when imaging to a fixed depth for certain regions in the Galaxy. For reasons of performing PSF photometry at the position of the brightest star in the difference images, we move the pixel coordinates of the star associated with the largest $\mathcal{F}$ value to the centre of the image.

(iv) For each star, the normalised Gaussian profile generated in (i) is placed at the appropriate pixel coordinates, and scaled by the star flux from (iii).

(v) Finally, a constant background level $S_R$ is drawn from the continuous distribution, $U \sim (10, 1000)$, is added to the image.

(vi) It is common practice to create a high signal-to-noise reference frame by either stacking images or increasing the exposure time. To simulate this, we generate a 'noise map', $\sigma_{R, ij}$ to apply to $R_{\textrm{noiseless}, ij}$ with 10 times less pixel variance as is applied to the target image. This can be achieved by scaling the uncertainties by a factor of $10^{-1/2} \sim 0.316$. Adopting the usual CCD noise model (Equation \ref{eq:CCDNM}) with $\sigma_0 = 5$ (ADU), $G=1$ $(\textrm{e}^- / \textrm{ADU})$ and $F_{ij}=1$, we compute the reference frame pixel uncertainties as
\begin{equation}
    \sigma_{R, ij} = 10^{-1/2}\sqrt{\sigma_{0}^{2} + R_{\textrm{noiseless}, ij}}\;.
    \label{eq:R_noisemap}
\end{equation}

(vii) A $142 \times 142$ pixel image, $W_{ij}$, with values drawn from a standard normal distribution, $\mathcal{N}(0, 1)$, was also generated, and the noisy reference, $R_{\textrm{noisy}, ij}$, is formed as
\begin{equation}
    R_{\textrm{noisy}, ij} = R_{\textrm{noiseless}, ij} + W_{ij}\sigma_{R, ij} \;.
    \label{eq:R_noisy}
\end{equation}
For each $R_{\textrm{noisy}, ij}$ we then generate a corresponding target image.

(viii) As with the reference images, we choose a symmetrical Gaussian PSF for the target images, paramaterised by $\phi_I$. As the convolution of a Gaussian with a Gaussian is another Gaussian, the kernel is itself a Gaussian, and we draw the corresponding kernel width $\phi_K$ from $U \sim (0.5, 2.5)$. We can then compute the width of the PSF in the target image,

\begin{equation}
    \phi_{I}^{2} = \phi_{R}^{2} + \phi_{K}^{2} \;,
    \label{eq:truetargetPSF}
\end{equation}

and repeat steps (iv) - (v), using $\phi_I$ in place of $\phi_R$ and $S_I$ in place of $S_R$ to generate $I_{\textrm{noiseless}, ij}$. Additionally, we also apply sub-pixel shifts to the stellar fractional pixel positions along both the x- and y-axis, with the $\Delta$x and $\Delta$y shifts each drawn separately for each simulation from $U \sim (-0.5, 0.5)$.

(ix) Similar to (vi), we then generate the noise map

\begin{equation}
    \sigma_{I, ij} = \sqrt{\sigma_{0}^{2} + I_{\textrm{noiseless}, ij}}\;,
    \label{eq:I_noisemap}
\end{equation}

(x) and then generate a $142 \times 142$ pixel image, $W_{ij}$, with values drawn from $\mathcal{N}(0, 1)$, and create $I_{\textrm{noisy}, ij}$,

\begin{equation}
    I_{\textrm{noisy}, ij} = I_{\textrm{noiseless}, ij} + W_{ij}\sigma_{I, ij} \;.
    \label{eq:I_noisy}
\end{equation}

The signal-to-noise ratio of the target image is computed as

\begin{equation}
    \text{SNR}_{I} = \frac{\sum_{ij}( I_{\textrm{noiseless}, ij} - S_I)}{\sqrt{\sum_{ij}\sigma_{I, ij}^2}}
    \label{eq:SNR_I}\;.
\end{equation}

\subsection{Performance Metrics}\label{sec:metrics}

We assess the fit quality and photometric accuracy of the kernel and background solutions with the following performance metrics. The derivations of these metrics are based on Section 5.2 of B16. For easy reference, the definitions and equation numbers for all metrics are listed in Table \ref{tab:metrics}. 

(i) Model error. The mean squared error (MSE) assesses how well the inferred model image, $M_{ij}$, matches the true target image, $I_{ij, \textrm{noiseless}}$,
\begin{equation}
    \textrm{MSE} = \frac{1}{N_{\textrm{data}}}\sum_{ij}(M_{ij}-I_{\textrm{noiseless}, ij})^2 \;.
    \label{eq:MSE}
\end{equation}
The smallest values of MSE indicate the best fit in terms of model error.

As noted in \citet{bramich2015difference}, systematic errors in the photometric scale factor, $P$, can badly influence photometric accuracy. Given this, we also assess the inferred $P$ and $B_0$ parameters, whose true values are equal to 1 and 0 respectively in these simulations.

(ii) Fit quality. The mean fit bias (MFB) and mean fit variance (MFV) are measures of the bias and excess variance in $M_ij$, and are given as
\begin{equation}
     \textrm{MFB} = \frac{1}{N_{\textrm{data}}}\sum_{ij}\frac{I_{\textrm{noisy}, ij} - M_{ij}}{\sigma_{I, ij}}\;,
     \label{eq:MFB}
\end{equation}
\begin{equation}
  \textrm{MFV} = \frac{1}{N_{\textrm{data}}-1}\sum_{ij}\left(\frac{I_{\textrm{noisy}, ij} - M_{ij}}{\sigma_{I, ij}}-\textrm{MFB} \right)^2\;.
  \label{eq:MFV}
\end{equation}
Kernel and background solutions with a MFB close to 0, and a MFV close to unity are measured to have a good fit quality.


(iii) Photometric accuracy. To assess the photometric accuracy of the solution, we perform PSF fitting photometry at the position of the brightest star in the difference image. We generate a normalised PSF object parameterised by $\phi_R$, centered at the position of the brightest star, and convolve this with the kernel solution. This is renormalised, giving us a normalised PSF object to fit to the difference image. The true target image PSF width is known by Equation \ref{eq:truetargetPSF}, and we set the size of the renormalised PSF object `stamp' to be $9\phi_I \times 9\phi_I$ pixels large. We then fit this PSF stamp to an equally sized cutout from the difference image (centered at the position of brightest star) with a scaling factor $\mathcal{F}_{\textrm{diff}}$ and an additive constant, weighted with the known pixel variances in the target image, $\sigma_{I, ij}^2$. The fitted $\mathcal{F}_{\textrm{diff}}$ is then scaled to the photometric scale of the reference image, giving $\mathcal{F}_{\textrm{measured}} = \mathcal{F}_{\textrm{diff}}/P$.

The theoretical minimum variance of $\mathcal{F}_{\textrm{measured}}$ for a PSF-fitting procedure which scales a PSF model to a stamp with pixel indices $rs$ is
\begin{equation}
    \sigma_{\textrm{min}}^{2} = \frac{1}{P_{\textrm{true}}^2}\left(\sum_{rs} \frac{\mathcal{P}_{I, rs}^2}{\sigma_{I, rs}^2}  \right)^{-1} \;,
    \label{eq:minvar}
\end{equation}
where $P_{\textrm{true}}$ is the true photometric scale factor (equal to 1 in our simulations) and $\mathcal{P}_{I, rs}$ is the true PSF of the brightest star in the target image i.e. a normalised Gaussian with standard deviation $\phi_I$.

As there are no variable sources, over $N_{\textrm{sim}}$ simulations of accurate kernel and background solutions we should expect a distribution of $\mathcal{F}_{\textrm{measured}}/\sigma_{\textrm{min}}$ values with mean 0 and a variance of unity. The mean photometric bias (MPB) and mean photometric variance (MPV) over $N_{\textrm{sim}}$ simulations, each indexed by $k$, would be equal to the statistics

\begin{equation}
    \textrm{MPB} = \frac{1}{N_{\textrm{sim}}}\sum_k \frac{\mathcal{F}_{\textrm{measured}, k}}{\sigma_{\textrm{min},k}}
    \label{eq:MPB}
\end{equation}

\begin{equation}
    \textrm{MPV} = \frac{1}{N_{\textrm{sim}} - 1}\sum_k\left(\frac{\mathcal{F}_{\textrm{measured}, k}}{\sigma_{\textrm{min},k}} - \textrm{MPB}\right)^2 \;.
    \label{eq:MPV}
\end{equation}

As noted in B16, although the MPV should be close to unity, it may have values less than this when the target image is overfitted, and/or when the model PSF fitted to the difference image (i.e. formed from the convolution of the reference image PSF with the inferred kernel) is different than the true PSF of the target image.

\begin{figure*}
\centering
\includegraphics[width=0.8\textwidth]{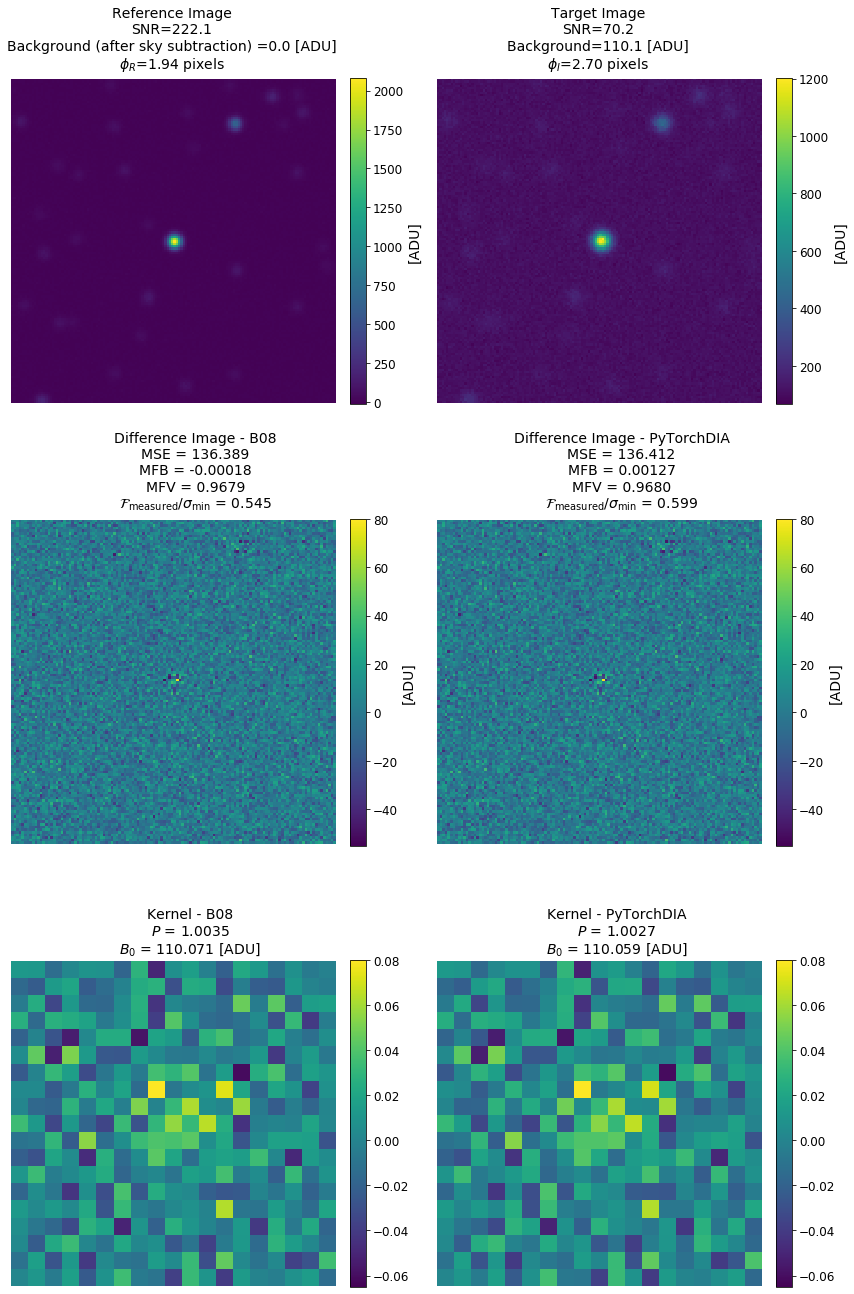}
\caption{Example images and fit quality metrics from the simulated image experiments. (\textit{Top row}) An example reference (left) and target (right) image pair generated in the simulation tests. (\textit{Middle row}) The subsequent difference images and fit metrics recovered by the B08 approach (left) and our PyTorch implementation (right). (\textit{Bottom row}) The corresponding convolution kernels and fit parameters for the B08 (left) and PyTorch (right) solutions. The median pixel value is subtracted from the reference image before fitting the kernel and differential background term, and so $B_0$ will not be equal to zero. The B08 and PyTorchDIA solutions are very similar.}
\label{fig:simulation_figs}
\end{figure*}

Figure \ref{fig:simulation_figs} shows an example $142 \times 142$ [pixels] reference and target image pair generated for these tests. The difference images and kernels corresponding to the B08 and PyTorchDIA solutions, annotated with the fit quality and photometric accuracy metrics, are shown below.

\begin{table}
    \centering
    \begin{tabular}{c|c|c}
       \hline \\
       Metric & Definition & Equation \\
       \hline \\
       $P$ & Photometric Scale Factor & \ref{eq:photometric_scale_factor} \\
       MSE & Mean Squared Error & \ref{eq:MSE}\\
       MFB & Mean Fit Bias & \ref{eq:MFB} \\
       MFV & Mean Fit Variance & \ref{eq:MFV} \\
       MPB & Mean Photometric Bias & \ref{eq:MPB} \\
       MPV & Mean Photometric Variance & \ref{eq:MPV} \\
    \end{tabular}
    \caption{The model accuracy, fit quality, and photometric performance metrics
    used to assess the DIA implementations in this work.}
    \label{tab:metrics}
\end{table}

\subsection{Simulated Image Test results}

In these tests, both B08 and PyTorchDIA attempt to fit the image model in Equation \ref{eq:model_image}, with an unregularised (square) $19\times19$ pixel kernel. The generated (square) reference and target images are each $142\times142$ pixels large, and so the number of target image pixels which enter the fit is $N_{\mathrm{data}} = (142 - 2\times9)^2 = 15376$. B16 used $141\times141$ large reference images and included 10201 target image pixels in their solution, and so the results presented here can be meaningfully compared against that prior work.

Following B16, we first divide the results of our 71569 simulation tests into 3 regimes by the signal-to-noise ratio (SNR) of the target image, $I$: (1) $8 < \text{SNR}_{I} < 40$; (2) $40 < \text{SNR}_{I} < 200$ and (3) $200 < \text{SNR}_{I} < 1000$. Each of these 3 categories is then divided into a 4 further categories by the sampling regime of the images: (i) $\phi_R$ > 1 and $\phi_K$ > 1, (ii)  $\phi_R$ > 1 and $\phi_K$ < 1, (iii)  $\phi_R$ < 1 and $\phi_K$ > 1, (iv) $\phi_R$ < 1 and $\phi_K$ < 1.

\begin{figure}
    \centering
    \includegraphics[width=\columnwidth]{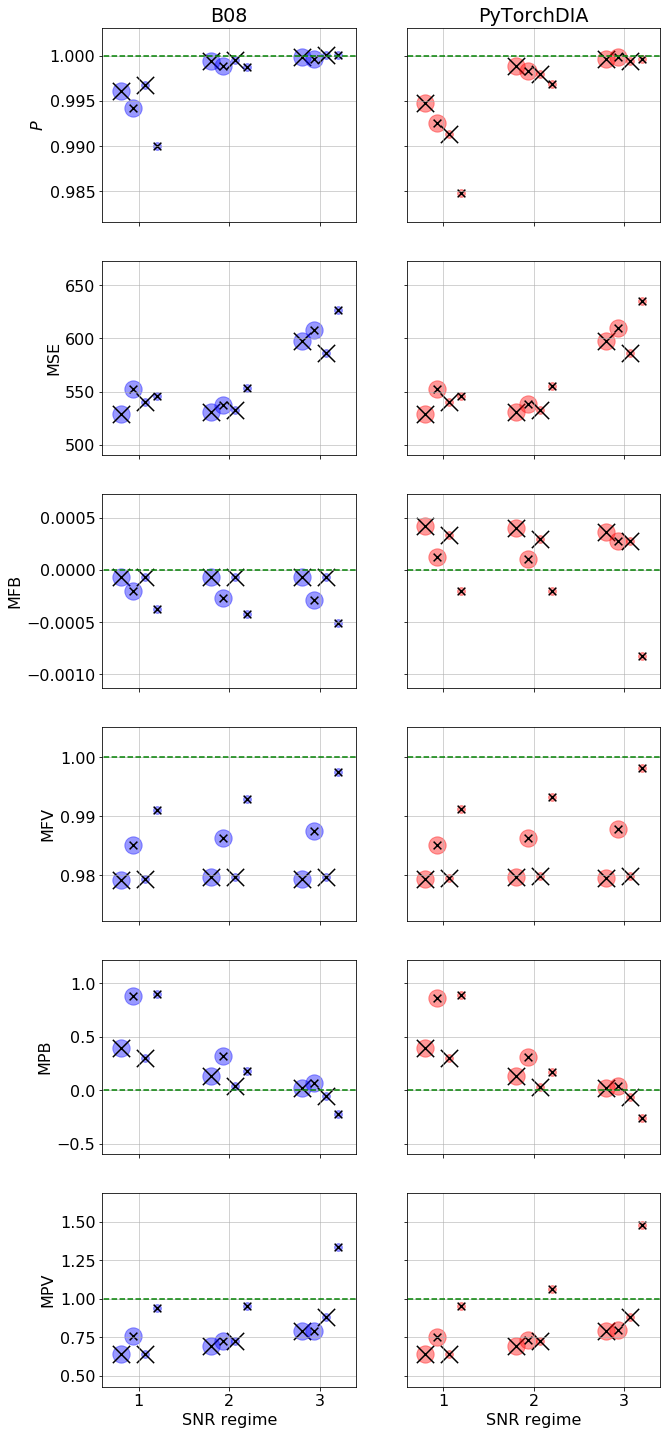}
    \caption{Fit quality and photometric accuracy metrics from the 71569 simulated image tests. Results for the B08 algorithm are in blue (left column), and the PyTorchDIA results are in red (right column). The signal-to-noise regime of the target image is shown on the x-axis, increasing from left to right. Metrics in each SNR regime are offset from each other for clarity. We use circular markers to denote the sampling regime of the reference image, and crosses to indicate the sampling regime of the kernel. A big circle or cross corresponds to an over-sampled reference image or kernel respectively, and a small circle or cross corresponds to an under-sampled reference image or kernel; there are therefore 4 possible combinations of marker for each SNR regime. The green dashed lines for each sub-plot pair represent the correct, `ideal' value for each metric.}
\label{fig:simulation_results}
\end{figure}

The distributions of the fit metrics are in general skewed, and so we report the median values as a robust estimate of the central value in Figure \ref{fig:simulation_results} for both the B08 and PyTorchDIA implementations. Each sub-plot pair in Figure \ref{fig:simulation_results} pertains to a single metric, with the B08 results in blue in the left column, and PyTorchDIA results in red in the right column. On the x-axes, we plot the SNR regime of the target image, categorised as above. There are 4 points in each SNR regime in each sub-plot, each of which corresponds to a different sampling regime, and we offset these from each other for clarity. We use circular markers to denote the sampling regime of the reference image, and crosses to indicate the sampling regime of the kernel. A big circle or cross corresponds to an over-sampled reference image or kernel respectively, and a small circle or cross corresponds to an under-sampled reference image or kernel. The green dashed lines for each sub-plot pair represent the `best' value for each metric. We also tabulate these results in Table \ref{tab:simulation_results_tab}, and include the 16th and 84th percentiles about the median of the distributions. In the bottom section of this table, we note the number of simulations which fall into each SNR and sampling regime category.

Across all the metrics, one of the biggest differences between the B08 and PyTorchDIA solutions are seen in the values for the photometric scale factor. B08 is more accurate in general, but differences between the median values in each SNR and sampling regime are typically small ($\sim 0.001$), and become negligible when the SNR is high. B16 showed that the accuracy of $P$ is strongly determined by the SNR of the target image, and in Table \ref{tab:simulation_results_tab}, we also see that the distribution of $P$ values about the median becomes tightly concentrated about unity as the SNR increases. Interestingly, for results in any given sampling regime, there is no clear similarity in trends between the two approaches with the target image SNR.

There are no large differences between B08 and PyTorchDIA in terms of MSE, with B08 better only at the level of the first decimal place. Looking at Table \ref{tab:simulation_results_tab}, B08 does however begin to slightly outperform PyTorchDIA in the highest SNR regimes when both the kernel and the reference frame are under-sampled.

As observed in B16, both approaches show overall gradually decreasing MFB as the SNR increases, although B08 shows a negative bias while PyTorchDIA is typically positively bias. PyTorchDIA again only appears to do noticeably worse than B08 in the highest SNR category when both the reference image and the kernel are undersampled.

The MFV values returned by B08 and PyTorchDIA are also very similar. The B08 MFV values are usually lower than those for PyTorchDIA, although all median values are less than unity. These results too are consistent with B16, who also found the unregularised kernel was prone to overfitting when the SNR of the target image is low.

The differences between B08 and PyTorchDIA in terms of MPB are small, with both showing the same trends with SNR and sampling regimes. As found by B16, with increasing SNR, the MPB greatly improves. That this is typically best when the kernel is oversampled, and worst when both the kernel and reference image are undersampled, is likely in part due to sampling issues associated with the PSF model which is scaled to the flux in the difference image. Encouragingly, with reference to the values in Table \ref{tab:simulation_results_tab}, we can see that in general, PyTorchDIA performs even better than B08. That the MPB metrics are similar despite the differences in MFB is likely due to our choice to simultaneously fit for a constant scalar offset in our PSF fitting procedure, which would correct for any net over-/underestimation in the background parameter of the image model.

The same trends in the MPV values with both SNR and sampling regimes can again be seen by the two approaches. With an increasing SNR, the MPV increases. This behaviour was also seen by B16, and it can be explained as reduced overfitting of the bright central stars from which this metric is computed. For both approaches, apart from some cases when both the kernel and reference image are undersampled, the MPV is always less than unity, and similar results were obtained by B16 (see Figure 6 therein). We experimented by performing an additional 21271 tests in which the photometry of the faintest star in the image was used to compute the MPV. In this instance, no MPV values were less than 1.1.

While being overall very similar, the PyTorchDIA MPV values are slightly higher than their B08 counterparts. Because B08 overfits more strongly than PyTorchDIA (see MFV values), we should expect the noise in the B08 difference images to be slightly more suppressed. Consequently, the photometry will tyically have a lower variance. The difference in MPV is greatest when both the kernel and reference image are undersampled. This may be associated with the worse MFB values for PyTorchDIA in this category, as while the additive constant in our PSF fitting procedure will be able to correct for inaccuracies in the differential background, it will do so at the cost of slightly increased variance in the photometry.

In conclusion, PyTorchDIA performs very similarly to the B08 algorithm in these tests on synthetic images, and it is encouraging to see that the major conclusions drawn from these tests are consistent with those in B16. These images have Gaussian noise contributions, and have no outliers, and it is therefore not surprising that the B08 algorithm does slightly better overall, since the linear least-squares formulation (within an iterative scheme) correctly describes the observation model, and this approach operates at twice the floating-point precision of PyTorchDIA. 

\renewcommand{\arraystretch}{1.45}
\begin{table*}
    \centering
    \begin{tabular}{cccccc}
    \hline
    Metric & SNR regime & $\phi_R > 1, \phi_K > 1$ & $\phi_R > 1, \phi_K < 1$ & $\phi_R < 1,\phi_K > 1$ & $\phi_R < 1, \phi_K < 1$ \\
    \hline
    $P$ (B08) & 1 & $0.9961_{-0.0358}^{+0.0335}$ & $0.9942_{-0.0380}^{+0.0316}$ & $0.9968_{-0.0323}^{+0.0273}$ & $0.9900_{-0.0342}^{+0.0279}$ \\
    $P$ (PyTorchDIA) & 1 & $0.9948_{-0.0360}^{+0.0335}$ & $0.9926_{-0.0376}^{+0.0320}$ & $0.9913_{-0.0328}^{+0.0277}$ & $0.9848_{-0.0368}^{+0.0285}$ \\
    $P$ (B08) & 2 & $0.9994_{-0.0125}^{+0.0125}$ & $0.9989_{-0.0134}^{+0.0124}$ & $0.9996_{-0.0108}^{+0.0101}$ & $0.9987_{-0.0111}^{+0.0103}$ \\
    $P$ (PyTorchDIA) & 2 & $0.9989_{-0.0126}^{+0.0124}$ & $0.9983_{-0.0135}^{+0.0128}$ & $0.9980_{-0.0112}^{+0.0098}$ & $0.9969_{-0.0135}^{+0.0114}$ \\
    $P$ (B08) & 3 & $0.9998_{-0.0046}^{+0.0045}$ & $0.9997_{-0.0046}^{+0.0045}$ & $1.0001_{-0.0040}^{+0.0037}$ & $1.0001_{-0.0040}^{+0.0039}$ \\
    $P$ (PyTorchDIA) & 3 & $0.9996_{-0.0045}^{+0.0045}$ & $0.9999_{-0.0050}^{+0.0047}$ & $0.9995_{-0.0040}^{+0.0038}$ & $0.9996_{-0.0049}^{+0.0042}$ \\
    \hline
    MSE (B08) & 1 & $528.53_{-329.61}^{+328.79}$ & $552.62_{-349.03}^{+310.51}$ & $539.85_{-342.21}^{+318.94}$ & $545.34_{-340.47}^{+313.22}$ \\
    MSE (PyTorchDIA) & 1 & $528.54_{-329.60}^{+328.80}$ & $552.63_{-349.03}^{+310.51}$ & $539.86_{-342.21}^{+318.95}$ & $545.42_{-340.54}^{+314.59}$ \\
    MSE (B08) & 2 & $530.88_{-330.54}^{+343.43}$ & $537.47_{-328.53}^{+337.46}$ & $532.74_{-337.18}^{+342.01}$ & $553.47_{-347.22}^{+336.57}$ \\
    MSE (PyTorchDIA) & 2 & $530.90_{-330.55}^{+343.43}$ & $537.92_{-328.88}^{+337.16}$ & $532.82_{-337.24}^{+341.94}$ & $555.51_{-346.81}^{+337.07}$ \\
    MSE (B08) & 3 & $597.17_{-362.18}^{+352.67}$ & $608.09_{-361.96}^{+361.21}$ & $585.85_{-356.65}^{+365.45}$ & $626.40_{-366.67}^{+415.64}$ \\
    MSE (PyTorchDIA) & 3 & $597.33_{-362.28}^{+352.64}$ & $610.09_{-363.08}^{+361.05}$ & $585.93_{-356.49}^{+365.41}$ & $634.81_{-371.31}^{+415.59}$ \\
    \hline
    MFB (B08) & 1 & $-0.0001_{-0.0001}^{+0.0000}$ & $-0.0002_{-0.0002}^{+0.0001}$ & $-0.0001_{-0.0001}^{+0.0000}$ & $-0.0004_{-0.0003}^{+0.0002}$ \\
    MFB (PyTorchDIA) & 1 & $0.0004_{-0.0003}^{+0.0004}$ & $0.0001_{-0.0003}^{+0.0003}$ & $0.0003_{-0.0004}^{+0.0004}$ & $-0.0002_{-0.0006}^{+0.0005}$ \\
    MFB (B08) & 2 & $-0.0001_{-0.0001}^{+0.0000}$ & $-0.0003_{-0.0002}^{+0.0001}$ & $-0.0001_{-0.0001}^{+0.0000}$ & $-0.0004_{-0.0004}^{+0.0002}$ \\
    MFB (PyTorchDIA) & 2 & $0.0004_{-0.0004}^{+0.0005}$ & $0.0001_{-0.0008}^{+0.0009}$ & $0.0003_{-0.0003}^{+0.0004}$ & $-0.0002_{-0.0060}^{+0.0044}$ \\
    MFB (B08) & 3 & $-0.0001_{-0.0001}^{+0.0001}$ & $-0.0003_{-0.0003}^{+0.0001}$ & $-0.0001_{-0.0002}^{+0.0001}$ & $-0.0005_{-0.0008}^{+0.0003}$ \\
    MFB (PyTorchDIA) & 3 & $0.0004_{-0.0007}^{+0.0007}$ & $0.0003_{-0.0013}^{+0.0018}$ & $0.0003_{-0.0004}^{+0.0005}$ & $-0.0008_{-0.0044}^{+0.0025}$ \\
    \hline
    MFV (B08) & 1 & $0.9792_{-0.0112}^{+0.0116}$ & $0.9851_{-0.0115}^{+0.0116}$ & $0.9794_{-0.0115}^{+0.0114}$ & $0.9911_{-0.0127}^{+0.0115}$ \\
    MFV (PyTorchDIA) & 1 & $0.9793_{-0.0112}^{+0.0116}$ & $0.9851_{-0.0115}^{+0.0117}$ & $0.9794_{-0.0115}^{+0.0114}$ & $0.9912_{-0.0127}^{+0.0116}$ \\
    MFV (B08) & 2 & $0.9796_{-0.0116}^{+0.0113}$ & $0.9862_{-0.0123}^{+0.0121}$ & $0.9797_{-0.0117}^{+0.0113}$ & $0.9930_{-0.0133}^{+0.0143}$ \\
    MFV (PyTorchDIA) & 2 & $0.9797_{-0.0116}^{+0.0113}$ & $0.9863_{-0.0123}^{+0.0123}$ & $0.9798_{-0.0117}^{+0.0113}$ & $0.9933_{-0.0134}^{+0.0147}$ \\
    MFV (B08) & 3 & $0.9794_{-0.0112}^{+0.0114}$ & $0.9875_{-0.0117}^{+0.0122}$ & $0.9797_{-0.0113}^{+0.0118}$ & $0.9974_{-0.0156}^{+0.0304}$ \\
    MFV (PyTorchDIA) & 3 & $0.9796_{-0.0112}^{+0.0114}$ & $0.9878_{-0.0118}^{+0.0128}$ & $0.9799_{-0.0114}^{+0.0118}$ & $0.9982_{-0.0161}^{+0.0384}$ \\
    \hline
    MPB (B08) & 1 & $0.401$ & $0.880$ & $0.300$ & $0.900$ \\
    MPB (PyTorchDIA) & 1 & $0.400$ & $0.862$ & $0.300$ & $0.887$ \\
    MPB (B08) & 2 & $0.138$ & $0.324$ & $0.038$ & $0.184$ \\
    MPB (PyTorchDIA) & 2 & $0.136$ & $0.313$ & $0.036$ & $0.176$ \\
    MPB (B08) & 3 & $0.027$ & $0.074$ & $-0.055$ & $-0.219$ \\
    MPB (PyTorchDIA) & 3 & $0.023$ & $0.046$ & $-0.059$ & $-0.257$ \\
    \hline
    MPV (B08) & 1 & $0.640$ & $0.760$ & $0.640$ & $0.937$ \\
    MPV (PyTorchDIA) & 1 & $0.639$ & $0.754$ & $0.640$ & $0.955$ \\
    MPV (B08) & 2 & $0.692$ & $0.728$ & $0.725$ & $0.951$ \\
    MPV (PyTorchDIA) & 2 & $0.693$ & $0.728$ & $0.726$ & $1.066$ \\
    MPV (B08) & 3 & $0.789$ & $0.788$ & $0.880$ & $1.339$ \\
    MPV (PyTorchDIA) & 3 & $0.789$ & $0.794$ & $0.883$ & $1.478$ \\
    \hline
    & \textbf{1} & 12332 &  4151 & 4078 & 1349 \\ 
    $N_{\textrm{Simulations}}$& \textbf{2} & 13224 & 4447 & 4387 & 1526 \\ 
    & \textbf{3} & 14673 &  4900 & 4829 & 1673 \\ 
    \hline
    \end{tabular}
    \caption{Fit quality and photometric accuracy metrics for the 71569 simulated image tests for both an implementation of the B08 algorithm and PyTorchDIA. We divide the results of these tests by the SNR regime of the target image and the sampling regimes of the reference image and convolution kernel. The number of simulations used for the computation of the metrics in each of the 12 possible SNR and sampling regime categories is shown in the bottom section.}
    \label{tab:simulation_results_tab}
\end{table*}

\subsection{No `algorithmic' bias}\label{sec:algobias}

As PyTorchDIA is a numerical algorithm, it is essential to know to what precision it can recover the correct solution if random noise has been removed from the data. By default, PyTorchDIA operates at F32, giving us 6-8 decimal digits of precision.

Following Section 3.1 of B08, we perform the following tests (i) - (iv) where known, noiseless transformations are applied to a reference image to generate a target image. The $142\times 142$ pixel reference image used in what follows is generated randomly, as outlined in Section \ref{sec:genimages}. We compute the fractional error, $\epsilon$, of $P$ and $B_0$, as $\epsilon_P = |(P - P_{\mathrm{True}}) / P_{\mathrm{True}}|$ and $\epsilon_{B_0} = |(B_0 - B_{0\mathrm{ ,True}}) / B_{0\mathrm{ ,True}}|$\footnote{As always, we background subtract the reference frame, so $B_0 > 0$.}.

(i) The simplest possible test we can perform is to difference an image against a copy of itself. The kernel should be exactly 1 at the centre and 0 everywhere else.
Encouragingly, PyTorchDIA is able to return the correct answer to within the theoretical convergence tolerance, with fractional errors of $\epsilon_P = 4 \times 10^{-7}$ and $\epsilon_{B_0} = 1 \times 10^{-8}$. The central kernel pixel was correct to a precision of 7 decimal digits.

(ii) Next, we convolve the reference with a Gaussian kernel of $\phi_K = 1.5$ pixels. Once again, PyTorchDIA returns this Gaussian kernel to within its convergence tolerance, with $\epsilon_P = 3 \times 10^{-7}$ and $\epsilon_{B_0} = 1 \times 10^{-8}.$

(iii) The target image is created by shifting the reference image by one pixel in each of the positive $x$ and $y$ spatial directions, without resampling. The corresponding kernel should be the identity kernel (central pixel value of 1 and 0 elsewhere) shifted by one pixel in each of the negative $l$ and $m$ kernel  coordinates. PyTorchDIA returns a kernel with a peak pixel value of unity precise to F32 machine precision, and  $\epsilon_P = 4 \times 10^{-7}$ and $\epsilon_{B_0} = 1 \times 10^{-8}.$

(iv) The reference image is shifted by half a pixel in each of the positive $x$ and $y$ spatial directions to create the target image, an operation that requires the resampling of the reference image. We use a bicubic spline resampling method\footnote{Specifically, we use the function scipy.ndimage.interpolation.shift, with no pre-filtering.}. PyTorchDIA performs very well, successfully returning the complicated kernel, with fractional errors on $P$ and $B_0$ of $\epsilon_P = 4 \times 10^{-7}$ and $\epsilon_{B_0} = 1 \times 10^{-8}.$

As PyTorchDIA is able to recover the true fit parameters correct to the data type's decimal digit precision, we conclude that there is no bias associated with our numerical algorithm.

\section{Real Image Tests}\label{sec:realimage_tests}

DIA is a particularly effective tool for measuring the flux and positions of variable sources in crowded fields (see the start of Section \ref{sec:introduction} for some examples). The MiNDSTEp consortium performs follow-up observations of microlensed targets towards the galactic bulge with the high frame-rate EMCCD cameras on the Danish 1.54m telescope (DK154), at ESO La Silla \citep{skottfelt2015two}. Short 0.1 second exposures are shift-and-stacked to eliminate tip tilt distortions, and recover scenes with $\sim 2-3$ times better resolution than conventional long integrations. The combined effects of the mount (the DK154 rests on three points) and the shift-and-add procedure produces irregular, triangular PSFs, with a sharp peak and diffuse halo. With a scale of 0.09 arcseconds per pixel, this high resolution EMCCD instrument explores a subset of the sampling regimes covered in the CCD simulations in Section \ref{sec:simulatedimagetests}.

\subsection{Data and reductions}

We use a sample of 251 $512\times512$ pixel images, each consisting of up to 3000 shift-and-stacked 0.1 s short exposures\footnote{During the shift-and-stack procedure, if a shift is above a certain threshold the frame is rejected, and so some stack consist of slightly less than 3000 images.} (i.e. each stack is equivalent to at most a 5 minute exposure), obtained with the `red' camera (approximately equivalent to a broad SDSS `z' filter \citealt{fukugita1996sloan}) on the Danish 1.54m Lucky Imaging instrument between 2019-07-17 and 2019-07-30. The observations are of the centre of the OGLE-III-BLG101 microlensing field, $(l,b) = (0.1331^{\circ}, -1.9643^{\circ})$. With a pixel scale of 0.09 arcseconds per pixel, the camera covers a field of view of $45 \times 45\; \text{arcseconds}^2$. All images are bias subtracted and flat corrected, and the master flats associated with each night of observing are used in the noise model for the following DIA performance tests (Equation \ref{eq:EMCCDNM}).

To create a high signal-to-noise reference, 13 sharp images acquired on 2019-07-20 were registered to the same pixel grid using bicubic spline resampling. The stacked reference frame was constructed by then summing the registered images, and dividing by 13. To measure the PSF of this stacked image, a (symmetric) Gaussian PSF model was independently fit to 5 bright, isolated stars, from which we found a median width of $\phi_R = 2.89$ pixels. The median $\phi_I$ of the 238 remaining target images was $3.83$ pixels, with a maximum of $6.67$ pixels.

In preparation for assessing the photometric accuracy of the algorithms, we measured the fluxes and positions of the stars in the reference image. A total of 236 candidate sources were detected in the image. We avoided all stars near the reference image edges and those with a peak pixel flux less than $3 \times 10^{4}$ ADU. This gave us a reduced sample size of the 37 brightest stars. Due to processing time constraints (see below), this sample was further reduced to 30, approximately uniformly spaced stars. 
The light curves of these 30 stars can be used to compute the MPB and MPV metrics (see Section \ref{sec:realdatametrics}).

For both the reference and each of the 238 target images, a $142\times 142$ pixel region was cutout around the positions of each of the 30 stars. This approach avoids resampling the target images to align them with the reference, and prevents introducing correlated noise into the target images by interpolating between pixels.\footnote{We note that the noise in all shift-and-stacked images will be correlated to some degree due to the shift procedure.}

Cursory image model fits to these target image stamps identified kernels with a size of $7\phi_I \times 7\phi_I$ to give good results. For the tests in Section \ref{sec:simulatedimagetests}, a $9\phi_I \times 9\phi_I$ kernel corresponded to a $19\times 19$ pixel array. For these tests on real images, even though the kernel is set to be just $7\phi_I \times 7\phi_I$ large, due to the fine sampling of the LI images, the median size of the kernel in pixel-space for this data set is $27 \times 27$ pixels, with a maximum size of $47 \times 47$ pixels. For the B08 algorithm in particular -- which scales with the square of the number of $N_K$ kernel pixels -- processing times are much more expensive than in Section \ref{sec:simulatedimagetests}, and this is why we further reduced our sample of bright stars down from 37 to 30 to keep this investigation tractable.

This should give us a total of 7140 reference-target image pairs (i.e. $\sim 10 \%$ the size of the simulated image data set), but in some instances, for stars close to the borders of the target image, large shifts between this image and the reference meant that part of the $142 \times 142$ pixel region fell outside the image edge. Additionally, the measured signal-to-noise ratio of the target image in some instances exceeded $1000$, and so these were rejected to provide a more consistent comparison with the results from Section \ref{sec:simulatedimagetests}. After these cuts, we had a final sample of 6989 reference-target image pair stamps.

\subsection{Model performance metrics for real data}\label{sec:realdatametrics}

For this test, we compare the PyTorchDIA code implementing our robust loss function with $c=1.345$ (Equation \ref{eq:huber_nll}), against the B08 solution making use of sigma-clipping. We run the B08 approach for a total of 3 iterations, and clip $|\epsilon_{ij}|>5$ pixels on the third iteration only.

Although we are both limited to a smaller sample size and do not know the true noise properties or model parameters of the actual images and kernel solutions respectively, we can still use almost all of the metrics outlined in Section \ref{sec:metrics} to assess the accuracy of the implementations. As the true model image is unknown, we are unable to use the MSE metric to measure the model error. The photometric scale factor, $P$, while also unknown, can however be compared on a relative scale, which requires an estimate for $P_{\mathrm{true}}$. 

It was found that the inferred $P$ was correlated with the distance from the image centre along the x-axis. This is due to a change in the sky level along this axis associated with instrumental effects, which influences $P$ due to the anti-correlation between the photometric scale factor and $B_0$. Consequently, $P_{\mathrm{true}}$  for the entire image is not well represented by a single number (e.g. the median value from all 30 stamps). Given this, we divided the x-axis into 4 equally spaced regions, and computed 4 separate $P_{\mathrm{true}}$ values using the subset of stamps in each of these region. The measured $P$ from any stamp was then normalised by the $P_{\mathrm{true}}$ corresponding to the region it belonged to along the chip's x-axis. As B08 outperformed PyTorchDIA in terms of $P$ in Section \ref{sec:simulatedimagetests}, we use the $P$ values inferred by this approach to determine $P_{\mathrm{true}}$.

The remaining metrics require an accurate noise model to be meaningful. For these EMCCD images, we use Equation \ref{eq:EMCCDNM}, and substitute the $M_{ij}$ obtained on the third and final iteration of the B08 solution of each reference-target image pair to estimate the pixel uncertainties. These pixel uncertainties can then be used to calculate the MFB and MFV metrics, in place of $\sigma_{I, ij}$. Similarly, by substituting the target image for $I_{\textrm{noiseless}, ij}$ in Equation \ref{eq:SNR_I} we can use these pixel uncertainties to calculate the signal-to-noise ratio of the target image for each reference-target image pair. For the third iteration of the B08 approach, we also sigma-clip $|\epsilon_{ij}| > 5$ pixels to guard against variable sources or spurious pixels affecting the kernel solution. We use this associated bad pixel mask to omit these pixels from the calculation of the corresponding MFB and MFV metrics for both the B08 and PyTorchDIA solutions.

We perform PSF fitting photometry at the centre of the difference images obtained from each reference-target image pair (i.e. at the position of one of the 30 possible stars) to assess the photometric accuracy of the approach. We infer an empirical model PSF for the target image in the following way. For each reference-target image pair, we identify the peak pixel values of bright sources in the reference, and set all other pixel values to 0. We then use our PyTorchDIA implementation to infer the 'PSF' which convolves this scene of (approximate) delta-functions to the target image. That is, we are inferring an estimate of the PSF of the target image with the 'kernel' solution of the PyTorchDIA code. When inferring this empirical PSF, we weight all pixels equally, which gives the pixel values of the very brightest stars more weight relative to the other pixels in the image.

As in Section \ref{sec:metrics}, we use a small square stamp for the PSF fitting. We set the radius of this stamp (i.e. the half-width of the stamp) to 1.5 FWHMs of the target image (i.e. $\sim 1.5 \times 2.355 \times \phi_I$ pixels), rounded up to the nearest odd integer pixel. The PSF object is normalised, and scaled to the flux of the difference image. To guard against the influence of (possibly variable) neighbouring sources or spurious pixels in the stamp not captured by the PSF model, we scale the model to a star's differential flux under our robust loss function (Equation \ref{eq:huber_loss}), with fixed pixel uncertainties and $c=1.345$. 

\begin{figure*}
\centering
\includegraphics[width=0.8\textwidth]{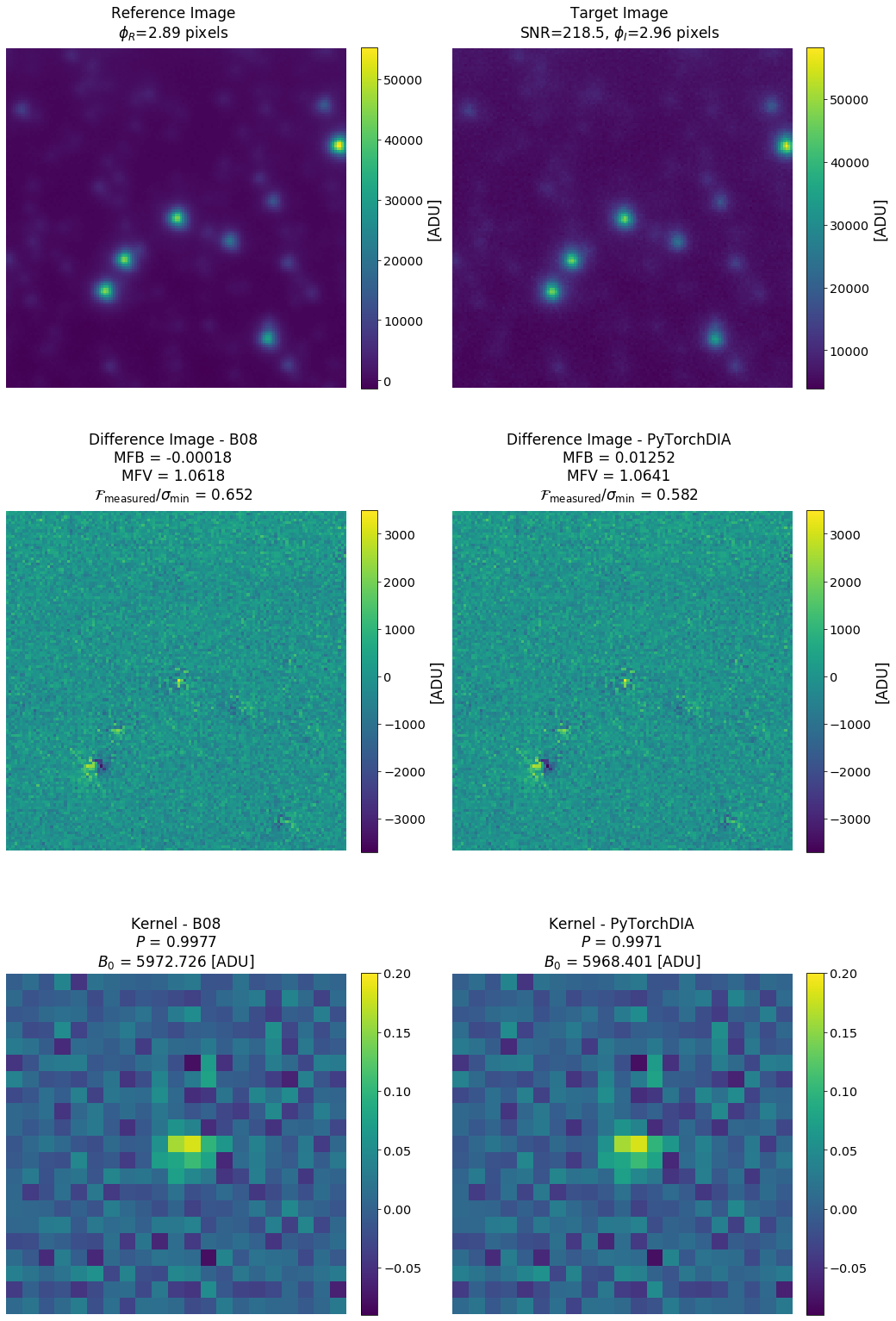}
\caption{Example reference-target image pair from the real image performance tests, in the same style as Figure \ref{fig:simulation_figs}. Note the unusual PSFs in the reference and target images, and the correspondingly irregular kernels. The target star with which the photometric accuracy was assessed is at the centre of the reference-target image pair.}
\label{fig:real_figs}
\end{figure*}

As with the MFB and MFV metrics, we adopt Equation \ref{eq:EMCCDNM} as the noise model, substituting the $M_{ij}$ values obtained on the final iteration of the B08 solution. In order to calculate $\sigma_{min}^{2}$ (Equation \ref{eq:minvar}), we substitute the normalised empirical PSF inferred by the PyTorchDIA code for $\mathcal{P}_{I, rs}$, and substitute the region-dependent $P_{\textrm{true}}$ inferred by the B08 implementation.
An example reference-target image pair and associated fit quality metrics for both the B08 and PyTorchDIA implementations are shown in Figure \ref{fig:real_figs}.

\subsection{Real Image Test results}

\begin{figure}
\centering
\includegraphics[width=\columnwidth]{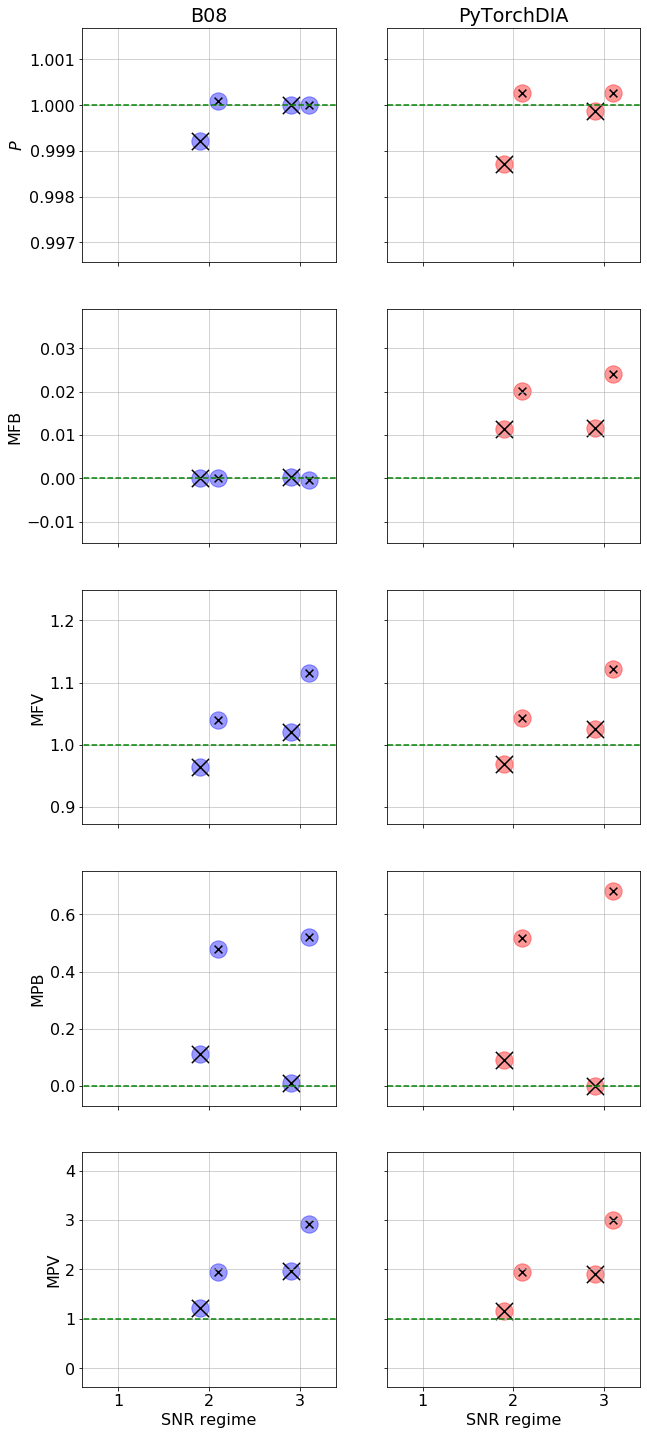}
\caption{Fit quality and photometric accuracy metrics from the 6989 real image tests. Results for the B08 algorithm are in blue (left column), and the PyTorchDIA results are in red (right column). The signal-to-noise regime of the target image is shown on the x-axis, increasing from left to right. We use circular markers to denote the sampling regime of the reference image, and crosses to indicate the sampling regime of the kernel. As the reference image is oversampled, all circular markers are large. A large cross corresponds to an oversampled kernel, and a small cross corresponds to an under-sampled kernel; there are therefore 2 possible combinations of marker for each SNR regime.No reference-target image pairs used to compute these metrics fell into the lowest SNR regime, and so that is left blank. The green dashed lines for each sub-plot pair represent the correct, `ideal' value for each metric.}
\label{fig:real_results}
\end{figure}

We again start our analysis by splitting our results into three signal-to-noise ratio (SNR) regimes by the signal-to-noise of the target image, $I$: (1) $8 < \text{SNR}_{I} < 40$;  (2) $40 < \text{SNR}_{I} < 200$ and (3) $200 < \text{SNR}_{I} < 1000$. We found that no reference-target image pairs correspond to the very lowest SNR regime, and so we are forced to restrict our analysis to  regimes (2) and (3). Also, unlike the simulated image tests, all 251 real images used here are well sampled. A majority 195 of the target images correspond to the case where the kernel is oversampled (i.e. $\phi_K > 1$), and for the other  images the kernel is undersampled (i.e. $\phi_K < 1$). We therefore further divide each signal-to-noise regime into two sampling regimes, dependent on whether the kernel is approximately over or undersampled. We plot the median metric values and the MPB and MPV for both B08 (blue) and PyTorchDIA (red), each split into the 4 possible sub-categories in Figure \ref{fig:real_results}.  As all images are oversampled, the circular markers are all large. The small and large crosses correspond to under- and oversampled kernels. The theoretical best value for each metric is plotted as a dashed green line. The median metrics and the 16th and 84th percentiles of their distributions are tabulated in Table \ref{tab:real_results}.

It was noted that some star differential light curves could have a noticeable non-zero offset from their reference frame flux level. In order to correct for this before grouping the normalised photometric residuals from across all light curves into each of the SNR and sampling regime categories, each star light curve was shifted such that its median photometric residual value was 0. Additionally, there was a small subset of outlying photometric residuals that may badly affect the MPB and MPV metrics. To remove the influence of these bad outliers in each category, we first calculate the median absolute deviation of the $\mathcal{F}_{\textrm{measured},k}/\sigma_{\textrm{min},k}$ values scaled to a standard deviation $\sigma_{\textrm{MAD}}$ as a robust estimate of the spread of the underlying distribution. This is defined as

\begin{equation}
    \sigma_{\textrm{MAD}} = \kappa \times
    \textrm{median}\left|\frac{\mathcal{F}_{\textrm{measured},k}}{\sigma_{\textrm{min},k}} - \textrm{median}\left(\frac{\mathcal{F}_{\textrm{measured}}}{{\sigma_{\textrm{min}}}}\right)\right| \;,
    \label{eq:sigma_MAD}
\end{equation}

where $\kappa = 1.4826$ for approximately normally distributed data. Before calculating the MPB and MPV metrics shown in Figure \ref{fig:real_results} and Table \ref{tab:real_results}, we have removed all photometric residuals with absolute values more than $4.5\sigma_{\textrm{MAD}}$ from $\textrm{median}(\mathcal{F}_{\textrm{measured}}/\sigma_{\textrm{min}})$ for each SNR and sampling regime category. Note that the number of outlying points may differ for the B08 and PyTorchDIA solutions.

We tabulate the number of reference-target image pairs in each category in Table \ref{tab:real_results}. Due to the sigma-clipping used before computing the MPB and MPV, we write the number of image pairs in each category used to compute these particular metrics as a fraction of the total number of image pairs. All image pairs in each category were used to compute all the other metrics, as these are robust to outliers.

We can again see that the PyTorchDIA and B08 approaches are broadly consistent, with both DIA implementations returning very similar metric values. Across all SNR and sampling regimes, only the median MFB values show consistently large differences. PyTorchDIA exhibited notably larger MFB values in the tests on simulated images, and here too, the B08 approach performs better with respect to fit bias. The $P$ values for B08 appear to be superior, but recall that these are normalised to the median $P$ values obtained from each x-axis sub-region by the B08 algorithm, so we should expect the B08 $P$ values to be closer to unity. For both implementations, as seen in the tests in Section \ref{sec:simulatedimagetests} and similar experiments in B16, the accuracy with which $P$ can be determined clearly improves with increasing SNR.

Similar trends with both MFV and MPB are seen from both approaches. There is a relative paucity of results corresponding to an undersampled kernel -- particularly for SNR category 2 -- but for the richly populated categories corresponding to oversampled kernels, both of these metrics improve with increasing SNR, as also observed in the simulated tests.

The MPV metrics for both approaches are noticeably larger here than in the simulations. However, we again see that the photometric variance goes up with the SNR of the target image, and that both approaches are overall very similar. Unlike the simulations, there is no guarantee that the 30 selected stars are the brightest in their respective stamps, and each stamp is well populated with stars in this crowded field (see top panel of Figure \ref{fig:real_figs}). We should not then be overfitting the target stars, and so MPVs greater than unity are expected, since the fractional contribution of flux of the target to the entire stamp will, in general, be much less than unity (see e.g. Figure 10 of B16). Also, of the 43 images corresponding to an undersampled kernel, 23 of these images have a PSF sharper than the stacked reference. In these instances, the model image will inevitably fail to match the PSFs -- as it would in fact have to de-convolve, rather than convolve the reference -- which will result in bad artefacts at the position of sources in the difference image. Indeed, in each SNR regime, the MPV corresponding to the undersampled kernel is much worse than its oversampled counterpart. However, these effects are common to both approaches, and here, we stress that we simply want to highlight the similarity of the MPV metrics obtained by the B08 and PyTorchDIA implementations. Indeed, the PyTorchDIA metrics can only be compared in a relative sense to the B08 results, as the B08 solutions for $P$ and $\sigma_{ij}$ are used to compute these metrics for both approaches. As we do not know the ground-truth for these values, we cannot say whether B08 or PyTorchDIA is closest to the correct answer, but only how similar they are to each other.

In addition to the differences in floating-point precision that would have influenced the small differences in results in Section \ref{sec:simulatedimagetests}, the two approaches now assume two different noise distributions for the target image pixels. For this real, outlier populated data, PyTorchDIA optimises a robust scalar objective function, while the B08 approach uses an iterative sigma-clipping procedure, so we should expect this to contribute to differences between the metric values also. And of course, the sample sizes in each category are smaller than their simulated data counterparts (particularly so for the instance of $\phi_K < 1$ in SNR category 2) and therefore noisier, which could partly explain some of the small differences between the metric values for the two approaches.

In summary, similar trends in metrics with SNR and sampling regime categories are shown by both the B08 and PyTorchDIA implementations, and these trends resemble those found in our experiments on simulated images. This both validates that our simulated images are reasonable approximations of real data, and that our robust PyTorchDIA solution provides competitive photometric performance with the B08 algorithm when applied to real astronomical data sets.

\renewcommand{\arraystretch}{1.4}
\begin{table*}
    \centering
    \begin{tabular}{|p{3cm}|p{3cm}|p{3cm}|p{3cm}|}
    \hline
     Metric & \textbf{SNR regime} & $\mathbf{\phi_R > 1, \phi_K > 1}$ & $\mathbf{\phi_R > 1, \phi_K < 1}$\\
    \hline
    $P$ (B08) & 1 & - & - \\
    $P$ (PyTorchDIA) & 1 & - & - \\
    $P$ (B08) & 2 & $0.9992_{-0.0122}^{+0.0136}$ & $1.0000_{-0.0036}^{+0.0070}$ \\
    $P$ (PyTorchDIA) & 2 & $0.9987_{-0.0121}^{+0.0134}$ & $1.0004_{-0.0048}^{+0.0065}$ \\
    $P$ (B08) & 3 & $1.0000_{-0.0063}^{+0.0086}$ & $1.0000_{-0.0048}^{+0.0050}$ \\
    $P$ (PyTorchDIA) & 3 & $0.9999_{-0.0066}^{+0.0083}$ & $1.0003_{-0.0052}^{+0.0049}$ \\
    \hline
    MFB (B08) & 1 & - & - \\
    MFB (PyTorchDIA) & 1 & - & - \\
    MFB (B08) & 2 & $0.0001_{-0.0001}^{+0.0002}$ & $0.0000_{-0.0002}^{+0.0002}$ \\
    MFB (PyTorchDIA) & 2 & $0.0114_{-0.0036}^{+0.0038}$ & $0.0198_{-0.0054}^{+0.0082}$ \\
    MFB (B08) & 3 & $0.0003_{-0.0004}^{+0.0005}$ & $-0.0003_{-0.0009}^{+0.0007}$ \\
    MFB (PyTorchDIA) & 3 & $0.0116_{-0.0044}^{+0.0045}$ & $0.0244_{-0.0067}^{+0.0098}$ \\
    \hline
    MFV (B08) & 1 & - & - \\
    MFV (PyTorchDIA) & 1 & - & - \\
    MFV (B08) & 2 & $0.9649_{-0.0431}^{+0.0480}$ & $1.0392_{-0.0204}^{+0.0328}$ \\
    MFV (PyTorchDIA) & 2 & $0.9699_{-0.0402}^{+0.0477}$ & $1.0416_{-0.0209}^{+0.0331}$ \\
    MFV (B08) & 3 & $1.0200_{-0.0475}^{+0.0809}$ & $1.1168_{-0.0561}^{+0.0901}$ \\
    MFV (PyTorchDIA) & 3 & $1.0253_{-0.0473}^{+0.0827}$ & $1.1246_{-0.0602}^{+0.1212}$ \\
    \hline
    MPB (B08) & 1 & - & - \\
    MPB (PyTorchDIA) & 1 & - & - \\
    MPB (B08) & 2 & $0.112$ & $0.415$ \\
    MPB (PyTorchDIA) & 2 & $0.090$ & $0.456$ \\
    MPB (B08) & 3 & $0.008$ & $0.530$ \\
    MPB (PyTorchDIA) & 3 & $-0.001$ & $0.703$ \\
    \hline
    MPV (B08) & 1 & - & - \\
    MPV (PyTorchDIA) & 1 & - & - \\
    MPV (B08) & 2 & $1.220$ & $1.902$ \\
    MPV (PyTorchDIA) & 2 & $1.165$ & $1.892$ \\
    MPV (B08) & 3 & $1.979$ & $2.851$ \\
    MPV (PyTorchDIA) & 3 & $1.929$ & $2.945$ \\
    \hline
    $N_{\textrm{Stamps}}$ (B08) & \textbf{1} & - & -  \\ 
    $N_{\textrm{Stamps}}$ (PyTorchDIA) & \textbf{1} & - & -  \\ 
    $N_{\textrm{Stamps}}$ (B08) & \textbf{2} & 1981/1992 &  57/57 \\ 
    $N_{\textrm{Stamps}}$ (PyTorchDIA) & \textbf{2} & 1983/1992 &  57/57 \\ 
    $N_{\textrm{Stamps}}$ (B08) & \textbf{3} & 4060/4097 &  843/843 \\ 
    $N_{\textrm{Stamps}}$ (PyTorchDIA) & \textbf{3} & 4056/4097 & 842/843 \\
    \hline
    \end{tabular}
    \caption{Fit quality and photometric accuracy metrics for the 6989 real image tests for both an implementation of the B08 algorithm and PyTorchDIA, separated into the SNR and sampling regimes. The number of image stamps used to compute the metrics in each of the SNR and sampling regime categories is shown in the bottom section of the table. As outlying photometric residuals have been removed from some categories prior to computing the MPB and MPV metrics, we write the number of $N_{\textrm{Stamps}}$ values used in their computation as a fraction of the total number of stamp samples.}
    \label{tab:real_results}
\end{table*}

\section{Speed tests}\label{sec:speed_test}

In the speed tests in this Section, we compare the performance of our GPU-accelerated numerical implementation against a fast, analytical least-squares fit solution using compiled Cython code. The PyTorchDIA code is run on Google's Colab\footnote{\url{https://colab.research.google.com}} service, a free-to-use cloud-based computation environment. The instance used for these experiments was equipped with a NVIDIA Tesla K80 GPU, with 2496 CUDA cores (with `compute capability' 3.7), with up to 16280 Mb of GPU RAM free. The Cythonised B08 solution is run on an Intel(R) Xeon(R) CPU @ 2.30GHz, with $\sim 12.6$ Gb RAM available.

Having established the accuracy of our algorithm on both synthetic and real images in Sections \ref{sec:simulatedimagetests} and \ref{sec:realimage_tests}, we first provide a motivating test exploring the computational speed-up over the state of the art classical approach on a set of real DK154 microlensing observations in Section \ref{sec:emccdspeed}. We then use a pair of large synthetic images to formally explore the scaling of our algorithm with image and kernel size in Sections \ref{sec:synspeed} and \ref{sec:kernelscaling}.

\subsection{Real EMCCD images}\label{sec:emccdspeed}

For these tests, we use a typical microlensing data set obtained with the DK154's `red' camera during the 2019 MiNDSTEp observing season. The data set consists of 159 shift-and-added exposures, each consisting of as many as 1200 short 0.1 s exposures (i.e. $\sim 2$ minute integrations).

The sharpest image was chosen as the reference, with a $\phi_R = 1.82$ pixels. The median target image width was $\phi_I = 3.18$ pixels. All images were bias subtracted and flat corrected. The target images were then re-sampled using bicubic spline interpolation to align them with the reference.
We use the flat field used for the data reduction in our noise model (Equation \ref{eq:EMCCDNM}). For the B08 implementation, we perform 3 model iterations and employ sigma-clipping, masking pixels associated with $|\epsilon_{ij}| > 5$, to guard against variable sources (e.g. the gravitationally lensed target).  For our PyTorchDIA implementation, we minimise the negative log-likelihood corresponding to our robust loss function (Equation \ref{eq:huber_nll}). 

\begin{figure*}
  \centering
  \begin{minipage}{\columnwidth}
   \includegraphics[width=\columnwidth]{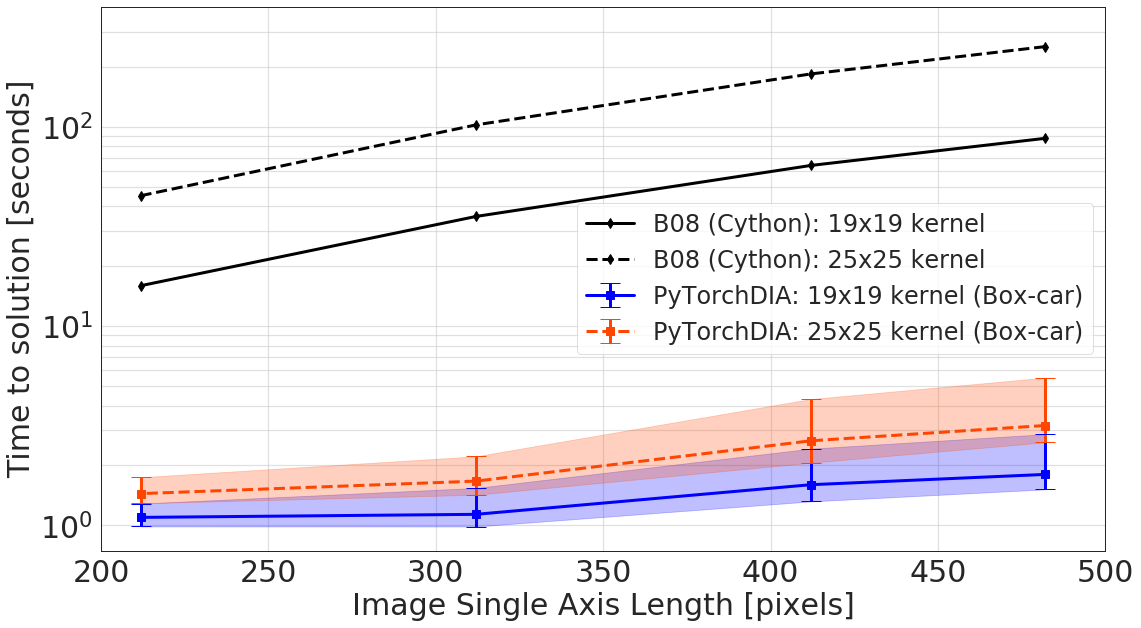}
  \end{minipage}
  \hfill
  \begin{minipage}{\columnwidth}
    \includegraphics[width=\columnwidth]{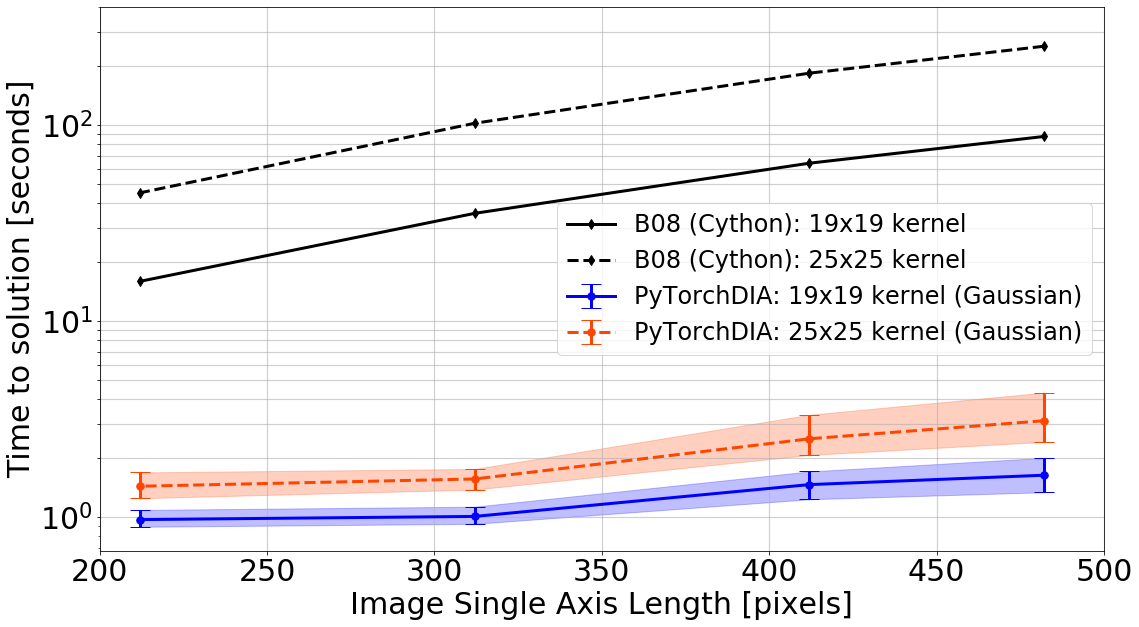}
  \end{minipage}
\caption{The time taken to solve for square kernels of different sizes against image single axis length (as a proxy for image size) for our PyTorchDIA implementation on the GPU and the B08 approach for images in a typical DK154 microlensing data set. The PyTorchDIA kernel pixels were initialised with a (left plot) `flat', box-car kernel and (right plot) a symmetric Gaussian with width estimated as $\phi_K = \sqrt{\phi_{I}^{2} - \phi_{R}^{2}}$.}
\label{fig:speedtest}
\end{figure*}

We test how the speed of our implementation scales with both image and kernel size by symmetrically cropping the borders of the images to different extents, and solving for both a $19\times 19$ and $25\times 25$ kernel. We measure the solutions times for each of the 158-large set of target images for our GPU-accelerated numerical algorithm, and plot the median solution time for a single image against the single axis length of the cropped square images (as a proxy for image size) in Figure \ref{fig:speedtest}, for two different choices of parameter initialisation. The `error' bars on the median image solution times are the 16th and 84th percentiles of the distribution of the 158 target image solution times. For comparison, we plot the time taken for 3 model iterations of the B08 analytical least-squares approach. We do not plot uncertainties for the B08 solution times. The total time of this analytic approach is dominated by the normal matrix construction, which is dependent only on the size of the kernel and images, and therefore effectively consistent across all 158 target images for each image and kernel size combination.

Unsurprisingly, the massive parallelisation inherent to the convolution computations in our implementation means it performs very well for even quite large image and kernel size combinations (see Section \ref{sec:synspeed} for a formal discussion of how the algorithm scales.). The median solution times for both the $19\times19$ and $25\times 25$ kernels are at least an order of magnitude faster than for the B08 approach. When scaled to data sets consisting of hundreds or thousands of images, this clearly provides substantial savings in processing times.

The two plots in Figure \ref{fig:speedtest} only differ in the choice of initialisation of the kernel pixels for PyTorchDIA. The plot on the left shows the solution times for a kernel initialised as a `flat', box-car which sums to 1. The right-hand plot shows the results for kernels initialised as symmetric Gaussians, normalised to sum to 1, with shape parameter estimated as $\phi_K = \sqrt{\phi_{I}^{2} - \phi_{R}^{2}}$. For these stacks of short EMCCD exposures, neither $\phi_I$ nor $\phi_R$ are particularly well approximated with a Gaussian, and the median solution times are at most only $\sim 1.1$ times faster. However, the spread of the distribution of solution times is clearly tighter when a Gaussian in used for the initialisation compared to the flat box-car.

\begin{figure}
    \centering
    \includegraphics[width=\columnwidth]{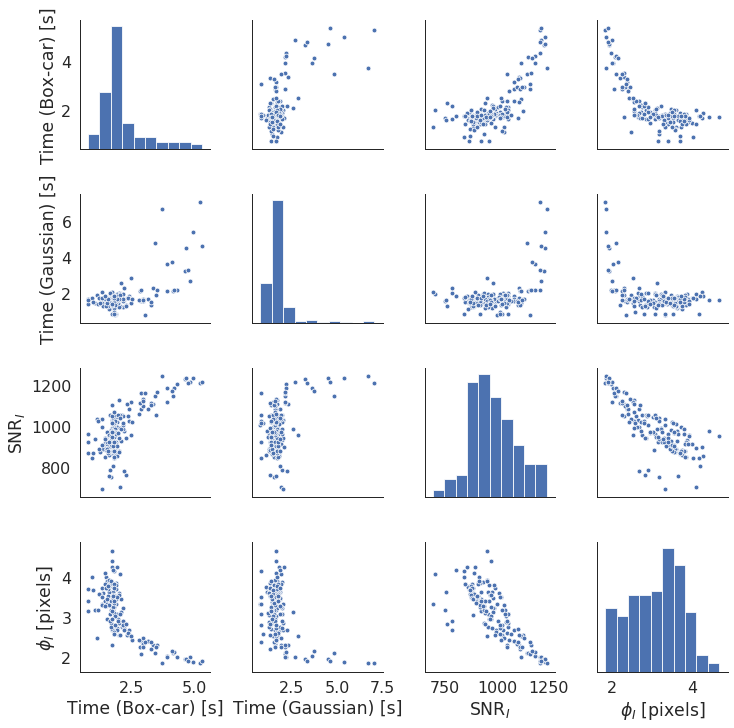}
    \caption{Pair plot showing the optimisation solution times for a $19\times 19$ kernel -- initialised as either a flat box-car or Gaussian -- on a $482\times 482$ large image, against the $\phi_I$ and SNR of the target image. The histograms of the distributions are shown on the diagonal.}
    \label{fig:speedtest_correlations}
\end{figure}

To gain some insight into the cause of the differences between the spread of values for the two different kernel initialisations (and therefore inform us of how to get the best performance out of PyTorchDIA) we plotted the solution times against the PSF width, $\phi_I$, and signal-to-noise ratio (SNR) of the target image for the set of $482\times 482$ images fit with a $19\times 19$ kernel as a pairplot in Figure \ref{fig:speedtest_correlations}. We quantified the correlation between solution times and these two image properties with the Spearman's rank correlation coefficient, $\rho$\footnote{This provides a measure of \textit{any} monotonic relation between two variables. The distributions (vs solution time) are clearly not linearly related, and so Pearson's correlation coefficient -- which assumes a linear relation -- is not suitable here.}. Both kernel initialisations result in solution times which are anticorrelated with $\phi_I$ ($\rho_{\phi_I, \mathrm{Box-car}}=-0.69$ and $\rho_{\phi_I, \mathrm{Gaussian}}=-0.29$) and correlated with the signal-to-noise ratio ($\rho_{\mathrm{SNR, Box-car}}=0.63$ and $\rho_{\mathrm{SNR, Gaussian}}=0.26$). We expect however that the cause of these trends is due only to $\phi_I$ and specifically, the \textit{difference} between the width of the PSFs in the target image and reference image. That the solution time is correlated with the signal-to-noise of the target image is due to the fact that the signal-to-noise is strongly anticorrelated with $\phi_I$.

The very strong anti-correlation between the solution time for the box-car kernel and $\phi_I$ is very likely due to this `flat' kernel surface being a much better initialisation for wide kernels. When the PSFs between the reference and the target images are very similar, the kernel is sharply peaked at the centre, and a flat kernel is far from the solution. Initialising with a Gaussian, although imperfect for the irregular PSFs associated with these images, appears to mitigate this problem to some degree, although the solution time still blows-up when the kernel is approximately under-sampled.

\subsection{Synthetic CCD images}\label{sec:synspeed}

Here, we test the performance of our implementation on synthetic images. We use our synthetic image generation procedure (see Section \ref{sec:genimages}) to first generate a large $4000 \times 4000$ pixel noiseless reference image. We set the logarithm of the stellar density per $100 \times 100$ pixels to be 1.5, we set $\phi_R = 1$ and the sky level at 100 ADU. Fluxes are assigned randomly as described in Section \ref{sec:genimages}. We set $\phi_K = 1.5$, and generate a target image with $\phi_I = \sqrt{\phi_{R}^{2} + \phi_{K}^{2}}$ in the same manner as above. We then randomly add noise to the images in way described in Section \ref{sec:genimages}, such that the target image has 10 times more pixel variance than the reference.

\begin{figure}
    \centering
    \includegraphics[width=\columnwidth, trim={1cm, 0, 1cm, 1cm}]{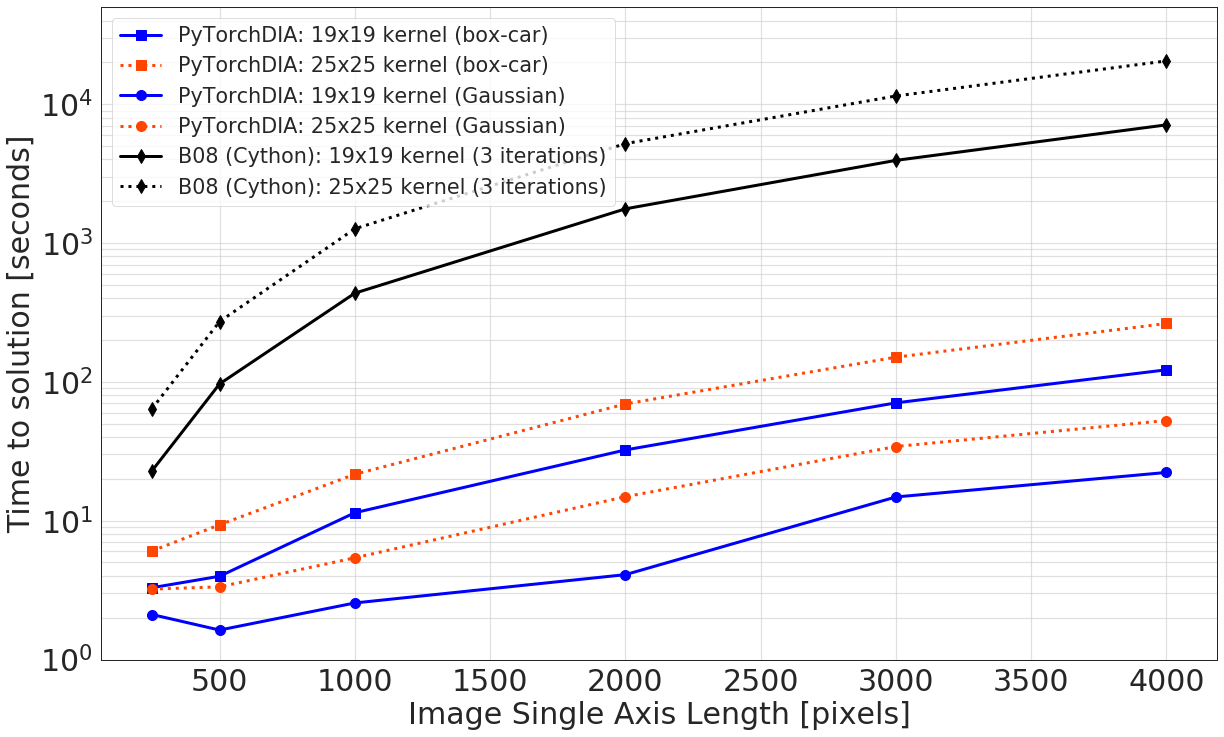}
    \caption{The time taken to solve for (square) kernels of different sizes against image single axis length (as a proxy for image size) for our PyTorchDIA implementation on the GPU and the B08 approach for a pair of square synthetic images, cropped to different sizes.}
    \label{fig:speedtest_synthetic}
\end{figure}

As there are no outlying pixels in this simulated pair of images, for the PyTorchDIA implementation, we minimise the Gaussian negative log-likelihood (Equation \ref{eq:loglikelihood}). The pair of $4000\times 4000$ images are symmetrically cropped to assess how the solution times scale with image size. For each kernel and image size combination (and choice of kernel initialisation), we plot the solution times in Figure \ref{fig:speedtest_synthetic}, for kernels initialised with either a flat box-car, or symmetrical Gaussian with width equal to $\phi_K = 1.5$.

As with the speed results in the prior Section, for larger images, the PyTorchDIA $19\times 19$ and $25\times 25$ kernel solution times are at least an order of magnitude faster than their B08 counterparts. In this case, since the correct convolution kernel really is a Gaussian, initialising the kernel as a Gaussian substantially reduces the number of optimisation steps required before convergence, with these solution times typically being $\sim 4-5$ faster than with a box-car initialisation for the kernel.
Formally, the B08 normal matrix construction time scales as $\mathcal{O}(n^2m^4)$, where $n$ and $m$ are the sizes of the (square) images and kernel respectively. The two black curves in both Figures \ref{fig:speedtest} and \ref{fig:speedtest_synthetic} follow this scaling (i.e. $\sim(25/19)^4$). Our optimisation procedure computes a direct convolution, which is predicted to scale as $\mathcal{O}(n^2m^2)$. The data points corresponding to the PyTorchDIA solution times in Figure \ref{fig:speedtest_synthetic} approximately obey this rule -- in both the separation of the two curves and the separation between data points on each curve -- although we should expect some variation due to differences in the number of optimisation steps required before converging and, potentially, cuDNN's choice of algorithm for the convolution computations, which can depend on both kernel and image size (see Section \ref{sec:kernelscaling}). In general, we found the number of optimisation steps required for convergence to slightly decrease with increasing image size. This makes sense, in that the effective information content of the image is greater the larger it is, and so the gradient steps are more accurate. This explains why for the smallest image size in Figure \ref{fig:speedtest_synthetic} the time to solve for some of the kernels takes longer than expected. Our solution time is then, not completely determined by the formal $\mathcal{O}(n^2m^2)$ scaling.

\subsection{cuDNN: Accelerating convolution computations on NVIDIA GPUs}\label{sec:kernelscaling}

One major use-case of PyTorch is for deep learning, in which CNNs, consisting of many small kernels, are used for feature detection in image recognition tasks \citep[for example,][]{lawrence1997face, krizhevsky2012imagenet}. The current research interest in deep learning applications has motivated device manufacturers such as NVIDIA to develop highly tuned GPU-accelerated libraries, specifically designed to improve the performance of common processing operations (e.g. forward and backward convolution) with these small kernels. By default, PyTorch makes use of NVIDIA's cuDNN library to accelerate convolutions on NVIDIA GPUs. This library has access to both deterministic and non-deterministic algorithms to compute convolutions, which are selected heuristically to accelerate computations\footnote{While not explored here, cuDNN also contains tools to benchmark convolution computations for a given kernel and image size combination. When processing many images, it performs tests to assess which cuDNN algorithm performs best on the first example, caches this information, and uses this best performing algorithm on all further images in the data set. Turning this feature towards astronomical data set reduction with PyTorchDIA will be explored in future work.}. As the sizes of useful kernels in DIA are themselves fairly small, it is expected that PyTorchDIA will also benefit from these tools. Indeed, we have enabled the use of non-deterministic cuDNN algorithms to accelerate the computation of convolution operations in all the results presented so far in this paper. In this subsection, we explore the benefit of this library in the context of DIA.

\begin{figure}
    \centering
    \includegraphics[width=\columnwidth]{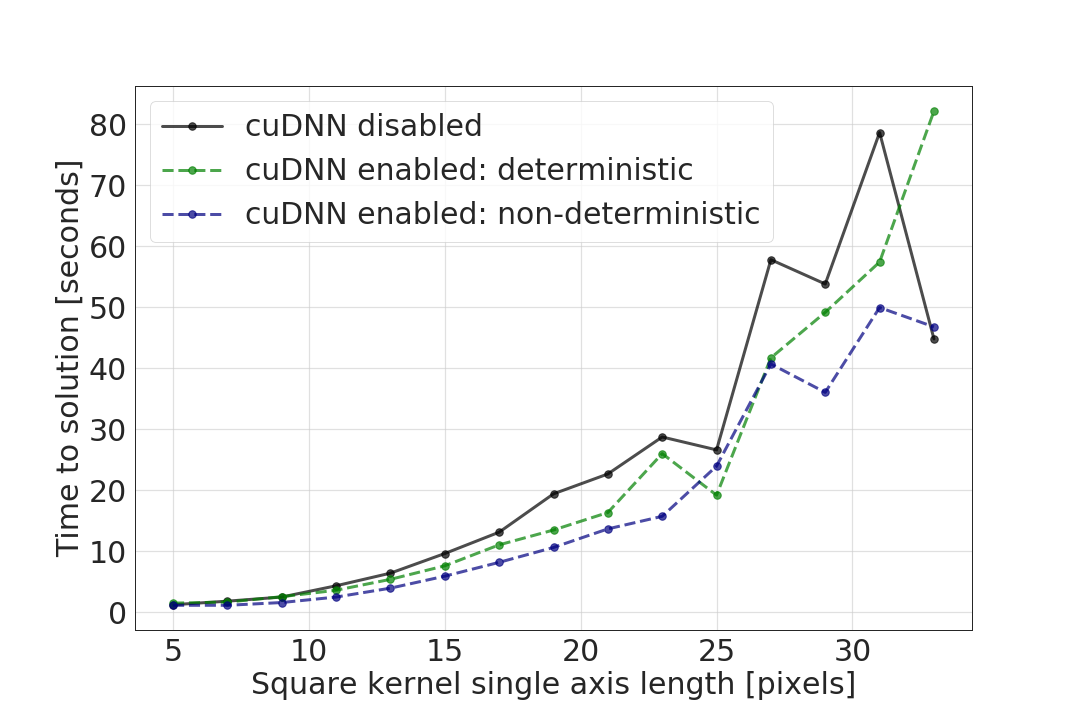}
    \caption{A comparison of the times taken to infer square kernels of different sizes with varying cuDNN settings for a given pair of $1000\times 1000$ pixel synthetic images.}
    \label{fig:kernel_scaling}
\end{figure}

For this test, we measure how for an image pair of fixed sizes, the time to infer the kernel scales with kernel size. We use the pair of synthetic $4000 \times 4000$ pixel images from the prior section, symmetrically cropped to smaller $1000 \times 1000$ images. We infer the associated convolution kernel (initialised with a box-car) by minimising the Gaussian negative log-likelihood for this pair of synthetic images, which contain no outlying pixels. The time to infer the kernel on the GPU for different cuDNN settings in shown in Figure \ref{fig:kernel_scaling}.

For all small kernels tested here, allowing cuDNN to use non-determinstic algorithms for the convolution computations results in slightly faster inference times. There is a clear change in behaviour for kernels larger than $23\times 23$ pixels, and while the non-deteriminsitic computations again seem to win overall, the scaling with kernel size is now far less predictable. This makes sense, as cuDNN is optimised for working with small kernels. Clearly however, for the larger kernels, small changes in kernel size can lead to sporadic changes in performance. This will most likely be due to changes in the choice of convolution algorithm implemented by the software for any given image-kernel size combination, although a full exploration of the different CUDA convolution algorithms, how they are chosen, and how they scale, is outside the scope of this work.

\section{Conclusions}\label{sec:conclusions}

We have presented a new algorithm for difference image analysis, where we model the target image as the output of a simple convolutional neural network. The kernel is treated as a discrete pixel convolutional filter, with weights which can be fit for efficiently within the PyTorch architecture. Specifically, we make use of automatic differentiation and GPU support to perform a lightning fast optimisation of the image model. We validate the fit quality and photometric accuracy of our implementation against its closest classical DIA analogue, with both simulated and real images. Our algorithm is simple to understand, and written entirely in standard Python packages, with an emphasis on accessibility to the wider astronomical community.

Its benefits over some traditional approaches can be summarised as follows:

\begin{itemize}
    \item \textit{Speed}: We exploit the massive parallelism inherent to the convolution operation by making use of highly-tuned GPU-accelerated deep learning libraries to perform efficient convolution computations. With a good choice of learning rate, the optimisation procedure converges rapidly, and our algorithm can perform DIA on astronomical data sets at least an order of magnitude faster than its classical analogue.
    \item \textit{Scalability}: For a pair of (square) images, each of size $n$, convolved with a (square) kernel of size $m$, our implementation approximately scales as $\mathcal{O}(n^2m^2)$, while the classical approach goes as $\mathcal{O}(n^2m^4)$.
    \item \textit{Flexibility}: In our optimisation framework, we can maximise the (correct) Gaussian likelihood suitable for conventional CCD/EMCCD astronomical exposures (where the Gaussian approximation of Poissonian photon noise is valid), without resorting to an iterative procedure of $\chi^2$ minimisation. This framework also allows us to relax the Gaussian noise assumption, and optimise robust scalar objective functions for images affected by outlying pixels. This provides a justifiable alternative to procedural sigma-clipping approaches.
    Further, we make use of automatic differentiation tools that free the user from having to manually recompute gradients if the paramaterisation of the model is changed, making experimentation by users straightforward.
\end{itemize}

The current main disadvantage to this approach is the use of 32-bit numerical precision to ensure the GPU-accelerated convolution computations are performed rapidly. Also, some engineering (e.g. choice of learning rates) is required by the researcher to get the best performance. And while the $\mathcal{O}(n^2m^2)$ scaling typically holds for a given image-pair and kernel combination, our approach is a non-linear (convex) optimisation, and so the solution times for our approach are not entirely deterministic. We also note that access to mid- to high-end GPUs can be an issue, and their use in both the gaming market and cryptocurrency mining has caused a surge in prices. Given this, Tensor Processing Units (TPUs) -- which now support floating point computations -- could be a viable alternative for some use-cases.

We highlight automatic mixed precision training, available since PyTorch 1.6.0, as an area to explore in future work to further accelerate the convolution computations. Given the impressive recent advances in general purpose GPU computing, in large part driven by the deep learning community, we are well positioned to take advantage of improvements to these tools. Lastly, we again stress that the application to DIA is just one possible example of astronomical image processing that can benefit from deep learning tools, as very many useful image models include a convolution operation.

All code used in this work can be found at \url{https://github.com/jah1994/PyTorchDIA}.

\section*{Acknowledgements}

We would like to thank the referee, Daniel Bramich, for his excellent review, which greatly improved the quality of this work. We also thank Keith Horne for his advice on specific details of our DIA implementation. We also extend a thank you to the MiNDSTEp consortium for allowing us use of imaging data from the 2019 observing season. Finally, the authors gratefully acknowledge funding from the Science and Technology Facilities Council of the United Kingdom.

This is a pre-copyedited, author-produced PDF of an article accepted for publication in Monthly Notices of the Royal Astronomical Society following peer review. The version of record -- \textit{Monthly Notices of the Royal Astronomical Society}, stab1114 -- is available online at: \url{https://doi.org/10.1093/mnras/stab1114}.

\section*{Data Availability}

The data underlying this article will be shared on reasonable request to the corresponding author.




\bibliographystyle{mnras}
\bibliography{PyTorchDIA} 





\bsp	
\label{lastpage}
\end{document}